\documentclass[a4paper,11pt]{article}
\usepackage{pos}
\usepackage{cleveref}
\usepackage[T1]{fontenc}
\usepackage{xspace}
\usepackage{wrapfig}
\usepackage{subcaption}
\usepackage{graphbox}
\usepackage{multicol}

\newcommand{\xmax}{\ensuremath{X_{\mathrm{max}}}\xspace}%

\title{Probing UHECR sources - Constraints from cosmic-ray measurements}
 \ShortTitle{Probing UHECR sources - Constraints from cosmic-ray measurements}

\author[a, b]{Teresa Bister}

\affiliation[a]{Nationaal instituut voor subatomaire fysica (NIKHEF), \\Science Park, Amsterdam, The Netherlands}

\affiliation[b]{Institute for Mathematics, Astrophysics and Particle Physics, Radboud University Nijmegen\\Nijmegen, The Netherlands}

\emailAdd{teresa.bister@ru.nl}

\abstract{Ultra-high-energy cosmic rays (UHECRs) are the most energetic particles known - and yet their origin is still an open question. However, with the precision and accumulated statistics of the Pierre Auger Observatory and the Telescope Array, in combination with advancements in theory and modeling - e.g. of the Galactic magnetic field - it is now possible to set solid constraints on the sources of UHECRs. The spectrum and composition measurements above the ankle can be well described by a population of extragalactic, homogeneously distributed sources emitting mostly intermediate-mass nuclei. Using additionally the observed anisotropy in the arrival directions, namely the large-scale dipole >8 EeV as well as smaller-scale warmspots at higher energies, even more powerful constraints on the density and distribution of sources can be placed. Yet, open questions remain - like the striking similarity of the sources that is necessary to describe the rather pure mass composition above the ankle, or the origin of the highest energy events whose tracked back directions point towards voids. The current findings and possible interpretation of UHECR data will be presented in this review.
}

\ConferenceLogo{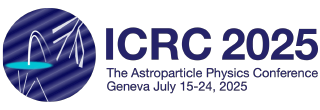}

\FullConference{39th International Cosmic Ray Conference (ICRC 2025)\\
 15–24 July 2025 \\
Geneva, Switzerland\\}

\begin{document}
\maketitle

\section{Introduction}
Although ultra-high-energy cosmic rays (UHECRs), nuclei with energies $>10^{18}\,\mathrm{eV}=1\,\mathrm{EeV}$, were first detected decades ago, their origin still remains an open question. What kind of powerful objects can accelerate particles to such energies, and how exactly the acceleration works, is still unclear. 
Through theoretical considerations, several constraints can be placed on UHECR sources, for example the Hillas criterion~\cite{hillas_origin_1984}, which gives a necessary geometrical condition on the size and magnetic field of the accelerator for the case of electromagnetic acceleration. Additionally, the sources must have sufficient power to generate particles of the observed energies~\cite{lovelace_dynamo_1976,waxman_cosmological_1995, blandford_acceleration_2000}.
Another criterion can be set by requiring the sources to supply the observed UHECR intensity, implying that the product of their number density and luminosity must be large enough. For more information see e.g. these reviews~\cite{anchordoqui_ultra-high-energy_2019, alves_batista_open_2019, matthews_particle_2020, globus_ultra_2025}.
Different types of possible steady or transient UHECR sources have been proposed and discussed in the literature, for example active galactic nuclei (AGNs), gamma-ray bursts (GRBs), starburst galaxies (SBGs), galaxy clusters, tidal disruption events (TDEs), binary neutron star (BNS) mergers, and more, see e.g.~\cite{alves_batista_open_2019, zhang_source_2025}. Several of these candidates fulfill the theoretical conditions, so definite conclusions regarding the sources of UHECRs must rely on measured data.

In the last years, a large increase in statistics has been achieved thanks to the two current biggest observatories for UHECRs measurements, the Pierre Auger Observatory~\cite{the_pierre_auger_collaboration_a_aab_et_al_pierre_2015} (\textit{Auger} in the following) in the southern hemisphere, and the Telescope Array~\cite{kawai_telescope_2008} (\textit{TA}) in the northern one. Due to their massive areas and extensive measurement times since 2004 and 2008, respectively, precise measurements of the energy spectrum and arrival directions of UHECRs are now available. Additionally, the mass composition can be constrained by means of the depth of shower maximum or the amount of muons in the shower~\cite{fitoussi_mass_2025}. 
Still, the interpretation of the data remains a delicate task. This is in part due to the influence of magnetic fields that deflect UHECRs on their way from the sources to Earth. Recently, significant progress has been made regarding models of the Galactic magnetic field (GMF)~\cite{unger_coherent_2024, korochkin_coherent_2025}, enabling far more precise predictions of the directional deflections of UHECRs. Nevertheless, the influence of the extragalactic magnetic field (EGMF) remains quite unknown (for a review see~\cite{durrer_cosmological_2013}). An upper bound on the field strength in voids from observations of the cosmic microwave background is $\mathcal{O}(1\,\mathrm{nG})$~\cite{ade_planck_2016}, but it is only applicable for a primordial origin of the EGMF. A comparable upper limit, independent of the EGMF origin, can however be placed using rotational measures~\cite{pshirkov_new_2016}. 
Next to the influence of the GMF and EGMF, also effects such as the interactions with background photon fields have to be considered when interpreting UHECR data. In recent years, models for the origin of UHECRs have become more and more sophisticated, allowing for better conclusions on UHECR sources, but also opening up even more questions.

In this review, recent progress regarding the interpretation of UHECR data will be discussed. For that, first the UHECR energy spectrum and composition data as well as their possible interpretation using models are presented in sec.~\ref{sec:CF}, and important common conclusions about UHECR sources will be identified. In sec.~\ref{sec:ADs}, further constraints from arrival direction data will be discussed, starting first with a summary of the effects of the Galactic magnetic field in sec.~\ref{sec:GMF}. Possible interpretations of the arrival direction data will be explored, moving from large-scale anisotropies in sec.~\ref{sec:LS} over intermediate scales in sec.~\ref{sec:IS} to the highest energy events in sec.~\ref{sec:high_E_events}. In the end, a conclusion and outlook to the future will be given in sec.~\ref{sec:conclusion}.

\section{Characteristics of UHECR sources - constraints from the measured spectrum and composition} \label{sec:CF}
One of the key measurement of UHECRs is the energy spectrum, where several features have  been significantly identified at different energies: the \textit{ankle} at around $10^{18.7}$~eV, the \textit{instep} or \textit{shoulder} at around $10^{19.1}$~eV, and a \textit{suppression} of the flux beyond $10^{19.7}$~eV. While the spectra measured in the southern hemisphere by Auger~\cite{the_pierre_auger_collaboration_a_aab_et_al_features_2020} and the northern hemisphere by TA~\cite{kim_energy_2025} agree about the general features and align well at lower energies when considering systematic uncertainties, there is some discrepancy at higher energies~\cite{tsunesada_measurement_2024} whose origin is still under debate.

The measurement of the depth of the shower maximum \xmax{} allows for conclusions on the primary mass of UHECRs. The Auger fluorescence detector data shows a transition from a mixture of lighter elements below the ankle towards continuously heavier mass at higher energies~\cite{the_pierre_auger_collaboration_a_aab_et_al_depth_2014, fitoussi_mass_2025}. The second moment of \xmax{} decreases continuously $\gtrsim5\,\mathrm{EeV}$, indicating that the amount of mixing between elements decreases with energy and the composition becomes progressively purer.
These trends were recently confirmed by surface detector data using a neural-network based analysis~\cite{pierre_auger_collaboration_inference_2025} which allowed for a 10-fold increase of statistics and an extension of the \xmax{} moments to higher energies. Also, it was shown that the Auger data is better described if the predicted \xmax{} by hadronic interaction models is shifted toward deeper values, implying an even heavier mass composition than previously inferred~\cite{the_pierre_auger_collaboration_testing_2024}.
While the TA data is often interpreted as light even above the ankle~\cite{kim_fractional_2025}, it was recently shown that it is compatible within uncertainties with both an all-proton as well as the heavier Auger mix interpretation~\cite{a_yushkov_for_the_pierre_auger_collaboration_and_the_telescope_array_collaboration_depth_2024}. The arrival directions measured at TA show a high level of isotropy implying a heavy mass composition, at least at the highest energies $\gtrsim100\,\mathrm{EeV}$~\cite{telescope_array_collaboration_isotropy_2024}. Soon, the mass composition at the highest energies will also be tested more directly using surface detector data as well by TA~\cite{prosekin_evaluation_2025}.

The energy spectrum and mass composition of UHECRs encodes valuable information about the source environment, acceleration, and propagation of UHECRs~\cite{gaisser_cosmic_2013, aloisio_transition_2012, allard_extragalactic_2012}. For example, the suppression at the highest energies was long believed to be due to the GZK effect~\cite{greisen_end_1966, zatsepin_upper_1966}, while it is now more probable to be at least partly due to the CR accelerators having reached their maximum energy around the GZK energy~\cite{the_pierre_auger_collaboration_a_aab_combined_2017}. The ankle was previously interpreted as the pair-production dip of an all-proton composition~\cite{berezinsky_astrophysical_2006}, which is now ruled out by Auger composition measurements~\cite{the_pierre_auger_collaboration_a_aab_et_al_depth_2014}.
By comparing a model including source injection, propagation, and detection to data, parameters of the model can be constrained to gain information about UHECR sources. Many such fits have been conducted in the last 15 years, see e.g.~\cite{hooper_heavy_2010, taylor_indications_2015, the_pierre_auger_collaboration_a_aab_combined_2017, unger_origin_2015, heinze_new_2019, mollerach_extragalactic_2020, d_bergman_for_the_telescope_array_collaboration_telescope_2021, the_pierre_auger_collaboration_a_abdul_halim_constraining_2023, the_pierre_auger_collaboration_a_abdul_halim_et_al_constraining_2024, abdul_halim_impact_2024}. 
Usually, these phenomenological fits start with an assumption about the emission of the sources, which is often parameterized as a power-law with a cutoff function $f_\mathrm{cut}(E, ...)$ that depends on the energy $E$ (plus additional parameters, see below):
\begin{equation} \label{eq:inj}
    Q_A(E) = Q_{0, A} \ \Big(\frac{E}{E_0} \Big)^{-\gamma} \ f_\mathrm{cut}(E, ...).
\end{equation}
Here, the spectral index $\gamma$ and the fractions $Q_{0,A}$ of some representative elements with charge number $Z$ and mass number $A$ are free model parameters that are adjusted to best describe the measured energy spectrum and $X_\mathrm{max}$ distributions through a maximum-likelihood fit. 

It is often assumed that UHECRs are accelerated electromagnetically, leading to a Peters cycle~\cite{peters_primary_1961} where the cutoff energy scales linearly with $Z$ so that $E_\mathrm{cut}:=Z \cdot R_\mathrm{cut}$. The rigidity cutoff $R_\mathrm{cut}$ is then another free model parameter. Often, a broken-exponential cutoff function $f_\mathrm{cut}(E / (Z \cdot R_\mathrm{cut})) = \exp(1- E / (Z \cdot R_\mathrm{cut}))$ that sets in only when the energy $E$ becomes bigger than $Z \cdot R_\mathrm{cut}$ is used instead of a simple exponential cutoff for better interpretability of the value of $\gamma$. Recently, parameterizations with more variability have been tried out successfully, such as $f_\mathrm{cut}(E / (Z \cdot R_\mathrm{cut}) = \text{sech}((E / (Z \cdot R_\mathrm{cut}))^\Delta)$, see e.g.~\cite{abdul_halim_impact_2024}. A value of $\Delta=2$ could for example stem from CR acceleration through magnetized turbulence~\cite{comisso_ultra-high-energy_2024}. An even more flexible parameterization $f_\mathrm{cut}(E, Z, A) \propto \exp(-E / (Z^\alpha A^\beta))$ has been investigated in~\cite{muzio_peters_2024}, which could account for energy losses in the source environment or beyond-standard-model physics. It was shown that the Peters cycle ($\alpha=1$ and $\beta=0$) may not be the optimal choice, but that signatures of different cutoff functions are not significantly differentiable from influences of additional source populations with current data.

Usually, one homogeneously distributed extragalactic source population is assumed to dominate above the ankle. Directly below the ankle, the mass composition is relatively light. Models where the transition from Galactic to extragalactic sources occurs at these energies can however be excluded by the low level of anisotropy contrary to expectations from light Galactic CRs at ankle energies~\cite{collaboration_constraints_2012}. Thus, current models often assume a second extragalactic population which consists of protons only and dominates below the ankle. This proton component can be either fitted freely~\cite{the_pierre_auger_collaboration_a_abdul_halim_et_al_constraining_2024, aloisio_ultra_2014}, or be assumed to be directly related to the high-energy population and generated through photo-disintegration in the source environment~\cite{unger_origin_2015, muzio_probing_2022, luce_observational_2022} (sometimes related to as the "UFA"-model~\cite{unger_origin_2015}). Explaining both components by a single population has the advantage that the smoothness of the energy spectrum, without any dips or abrupt breaks, is explained without finetuning of the normalizations of the two contributions.
Examples of works where UHECR (and neutrino) production in specific types of source candidates is modeled and compared to data are~\cite{baerwald_uhecr_2013, globus_complete_2015, globus_uhecr_2015, biehl_cosmic_2018, zhang_low-luminosity_2018, heinze_systematic_2020} for gamma-ray bursts (GRBs), \cite{rodrigues_active_2021, supanitsky_origin_2018, eichmann_ultra-high-energy_2018, eichmann_explaining_2022} for active galactic nuclei (AGNs), \cite{guepin_ultra-high-energy_2018, biehl_tidally_2018, plotko_ultra-high-energy_2024} for tidal disruption events (TDEs), \cite{condorelli_testing_2023} for starburst galaxies, \cite{fang_linking_2018} for black hole jets, and \cite{rossoni_investigating_2024} for binary neutron star (BNS) mergers. In addition to the secondary component from in-source interactions in the high-energy population, a subdominant intermediate~\cite{the_pierre_auger_collaboration_a_abdul_halim_et_al_constraining_2024} to heavy~\cite{unger_origin_2015} Galactic component is needed to explain spectrum and composition below the ankle.

\begin{figure}[ht]
\centering
\includegraphics[width=0.47\textwidth]{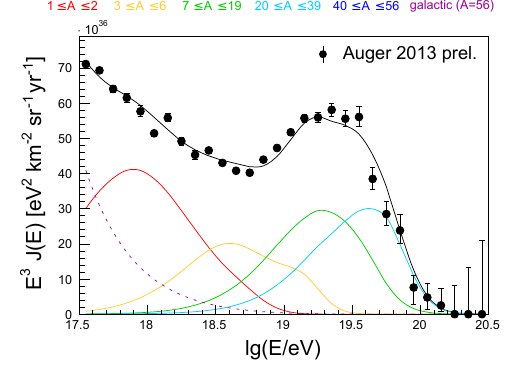}
\includegraphics[width=0.51\textwidth]{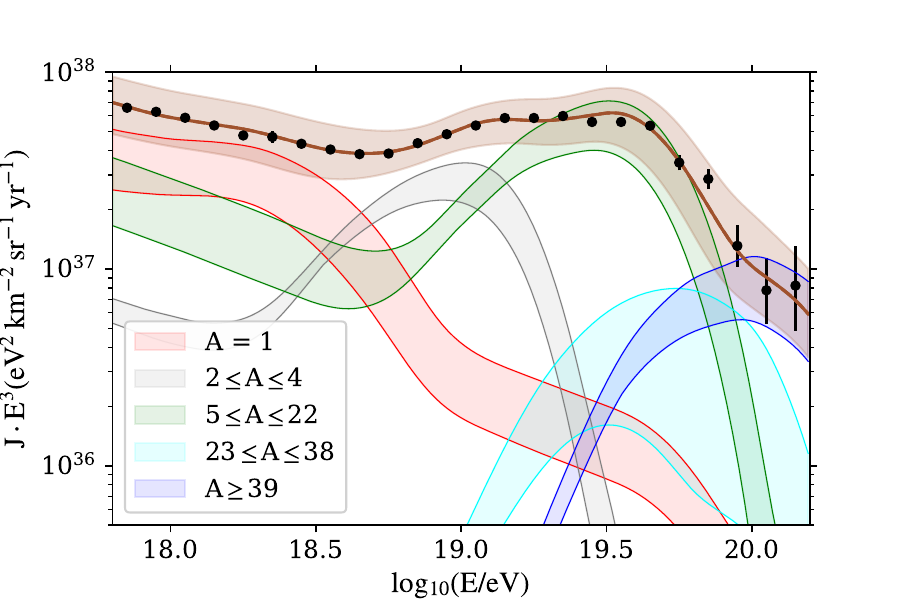}
\caption{Measured (black markers) and modeled (colors) energy spectrum at Earth. \textit{Left} for the "UFA"-model~\cite{unger_origin_2015} with one Galactic (dashed) and one extragalactic mixed source population considering also in-source interactions. \textit{Right} for a model with two extragalactic mixed populations without in-source interactions from~\cite{the_pierre_auger_collaboration_a_abdul_halim_constraining_2023}, where the bands indicate the size of the systematic uncertainties.}
\label{fig:spectrum}
\end{figure}

Another way to describe the UHECR data is by a second extragalactic population dominating directly below the ankle that also emits a mixed composition, in addition to the high-energy extragalactic component~\cite{the_pierre_auger_collaboration_a_abdul_halim_et_al_constraining_2024}. In that case, no Galactic component is needed for $E>5\,\mathrm{EeV}$. In Fig.~\ref{fig:spectrum}, the spectrum is shown, \textit{left} for the model with a secondary proton component from in-source interactions~\cite{unger_origin_2015}, and \textit{right} for the model with a second mixed extragalactic component~\cite{the_pierre_auger_collaboration_a_abdul_halim_constraining_2023}.

Even though all aforementioned models put emphasis on different details, several solid conclusions about the source population dominating above the ankle have emerged as common between fits to spectrum and composition data:
\begin{itemize}
    \vspace{-0.1cm}\item \textbf{A not too strong source evolution:} The redshift-evolution of the source population is often parameterized as $\propto(1+z)^m$ where $z$ is the redshift and $m$ a free model parameter. It has been shown that a strong source evolution with $m\gtrsim5$, where the sources are predominantly far-away, leads to an overproduction of low-energy secondaries and hence an overshoot of the spectrum below the ankle~\cite{the_pierre_auger_collaboration_a_abdul_halim_constraining_2023, the_pierre_auger_collaboration_a_abdul_halim_et_al_constraining_2024, alves_batista_cosmogenic_2019}. Additionally, strong source evolutions can lead to an overproduction of $\gamma$-rays not compatible with current limits~\cite{the_pierre_auger_collaboration_a_abdul_halim_constraining_2023, globus_probing_2017}, and can also overshoot the limits on cosmogenic neutrinos for the case of a secondary proton-producing population~\cite{ehlert_constraints_2023, muzio_prospects_2023}. A source evolution with $m=5$ is associated with intermediate-luminosity AGNs, and even stronger evolutions with high-luminosity AGNs~\cite{hasinger_luminosity-dependent_2005}, thus disfavoring both as the sole sources of UHECRs.
    \vspace{-0.2cm}\item \textbf{A hard emission spectrum \boldmath{$\gamma<1$} in combination with a mixed (often Nitrogen-dominated) composition:} This is necessary in order to describe the pronounced features of the spectrum, and the progressively heavier composition above the ankle in combination with the small mixing between elements indicated by the \xmax{} distributions~\cite{fitoussi_mass_2025}. As visible in Fig.~\ref{fig:spectrum}, each nuclear component only contributes to a small energy range. In those models, the instep is explained by the transition from Helium to Nitrogen and no additional flux contribution, e.g. from a local source, is necessary\footnote{Note that a scenario where the instep is generated by few foreground sources is disfavored also by the fact that the spectrum feature is consistently observed over the whole declination range covered by Auger~\cite{collaboration_energy_2025}.}.
    The major point of criticism about these models is that values for the spectral index $\gamma\ll2$ are unexpected from shock-acceleration. Note however that the spectrum given in eq.~\ref{eq:inj} relates to the emission leaving the source environment, so that the true acceleration spectrum can differ due to magnetic confinement and interactions in the source environment~\cite{unger_origin_2015}. Also, the inferred value of $\gamma$ is strongly influenced by the shape of the cutoff function~\cite{abdul_halim_impact_2024}, systematic uncertainties~\cite{the_pierre_auger_collaboration_a_abdul_halim_et_al_constraining_2024}, and the assumed source evolution (anti-correlation between $m$ and $\gamma$~\cite{alves_batista_cosmogenic_2019}). Additionally, the flux suppression of low-energy particles in the EGMF can have a substantial effect on $\gamma$~\cite{abdul_halim_impact_2024, mollerach_extragalactic_2020} - although extremely strong magnetic fields between the Milky Way and the first sources of around $B_\mathrm{rms}\approx10-200$~nG have to be assumed to get $\gamma$ in accordance with shock acceleration. A consequence of the hard Peters cycle source emission is that the predicted rigidity of UHECRs above the ankle stays relatively constant and with only a small spread at $R\approx4\pm3\,\mathrm{EV}$~\cite{bister_constraints_2024}.
    \vspace{-0.2cm}\item \textbf{Almost identical maximum source rigidities:} In~\cite{ehlert_curious_2023}, it was investigated to what extend it is justified to approximate all sources to be the same - while even within one source class variations between source luminosity, size, magnetic field etc. are expected. They built a model where the maximum source rigidity varies between candidates in the population $\propto R_\mathrm{cut}^{-\beta_\mathrm{pop}}$. It was found that values $\beta_\mathrm{pop}\gtrsim4$ are preferred, meaning that the population variance is surprisingly low within a factor of a few, and that UHECR sources are essentially "standard candles". This is necessary to explain the sharp features in the energy spectrum along with the small allowed mixing of the mass composition. In~\cite{Luce_ICRC_2025}, the small allowed variance of the maximum rigidity is confirmed, while they found that the spectral index allows for a larger variation.
    Explaining the unexpectedly narrow maximum rigidity range is currently one of the pressing question about UHECR sources. One proposed explanation is that BNS mergers are the sources of UHECRs, for which the small variance in NS masses driving the dynamo gravitationally could explain the similarity between maximum rigidities~\cite{farrar_binary_2025}.

\end{itemize}

Having established these findings, recent works have started to extend the combined fit framework, e.g. to constrain parameters of a possible EGMF~\cite{abdul_halim_impact_2024, bister_constraints_2024}, or the contribution of ultraheavy elements~\cite{zhang_ultraheavy_2024}. 
An exciting extension is also the inclusion of explicit source candidates in the model whose contribution to the flux can then be determined. Examples include FR0 radio galaxies~\cite{lundquist_combined_2025}, specific AGNs~\cite{eichmann_ultra-high-energy_2018, eichmann_explaining_2022}, catalogs of starburst galaxies, AGNs, or the nearby radio galaxy Centaurus A (hereafter Cen A)~\cite{the_pierre_auger_collaboration_a_abdul_halim_et_al_constraining_2024}, or sources with different number densities following the large-scale structure (LSS)~\cite{bister_constraints_2024}. As in most of these analyses the arrival directions have been used as an additional, powerful observable, these works will be discussed in more detail below.

In the future, the combined fit of spectrum and composition will become even more constraining thanks to the inclusion of composition data from the surface detectors of Auger and TA as described above. The strong increase in statistics and the addition of composition information at the highest energies will lead to stronger constraints on the source parameters and model variations.

\section{Constraints from cosmic-ray arrival directions} \label{sec:ADs}
On top of the constraints from the UHECR spectrum and composition measurements, the arrival directions offer the possibility to learn more especially about the spatial distribution of UHECR sources. As the arrival directions are heavily influenced by the Galactic magnetic field, its effect on the arrival flux of UHECRs will be described in the following section~\ref{sec:GMF}. Afterwards, the constraints from large- and intermediate-scale anisotropies as well as the highest energy events will be analyzed in sections~\ref{sec:LS}, \ref{sec:IS}, and \ref{sec:high_E_events}, respectively.

\subsection{Influence of the Galactic magnetic field} \label{sec:GMF}
\begin{figure}[htb]
\centering
\subfloat[Median deflection angle distribution over the sphere for different GMF models at rigidity $R=10\,\mathrm{EV}$, from~\cite{korochkin_uhecr_2025}. The median is around $30^\circ$ for both newest GMF models \texttt{UF23-base}~\cite{unger_coherent_2024} and \texttt{KST24}~\cite{korochkin_coherent_2025}, with a tail extending up to $\mathcal{O}(100^\circ)$.]{\label{fig:GMF_median_defl}\includegraphics[width=0.48\textwidth]{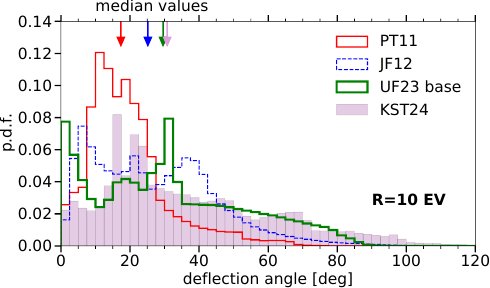}}\hfill
\subfloat[Flux magnification ($>1$, brownish) and demagnification ($<1$, greenish) in the UF23 GMF model suite at $R=5\,\mathrm{EV}$. The color bar shows regions where all 8 \texttt{UF23} models (with Planck-tuned JF12 random field) agree, while white denotes no agreement. From~\cite{bister_large-scale_2024}.]{\label{fig:GMF_demag}\includegraphics[width=0.48\textwidth]{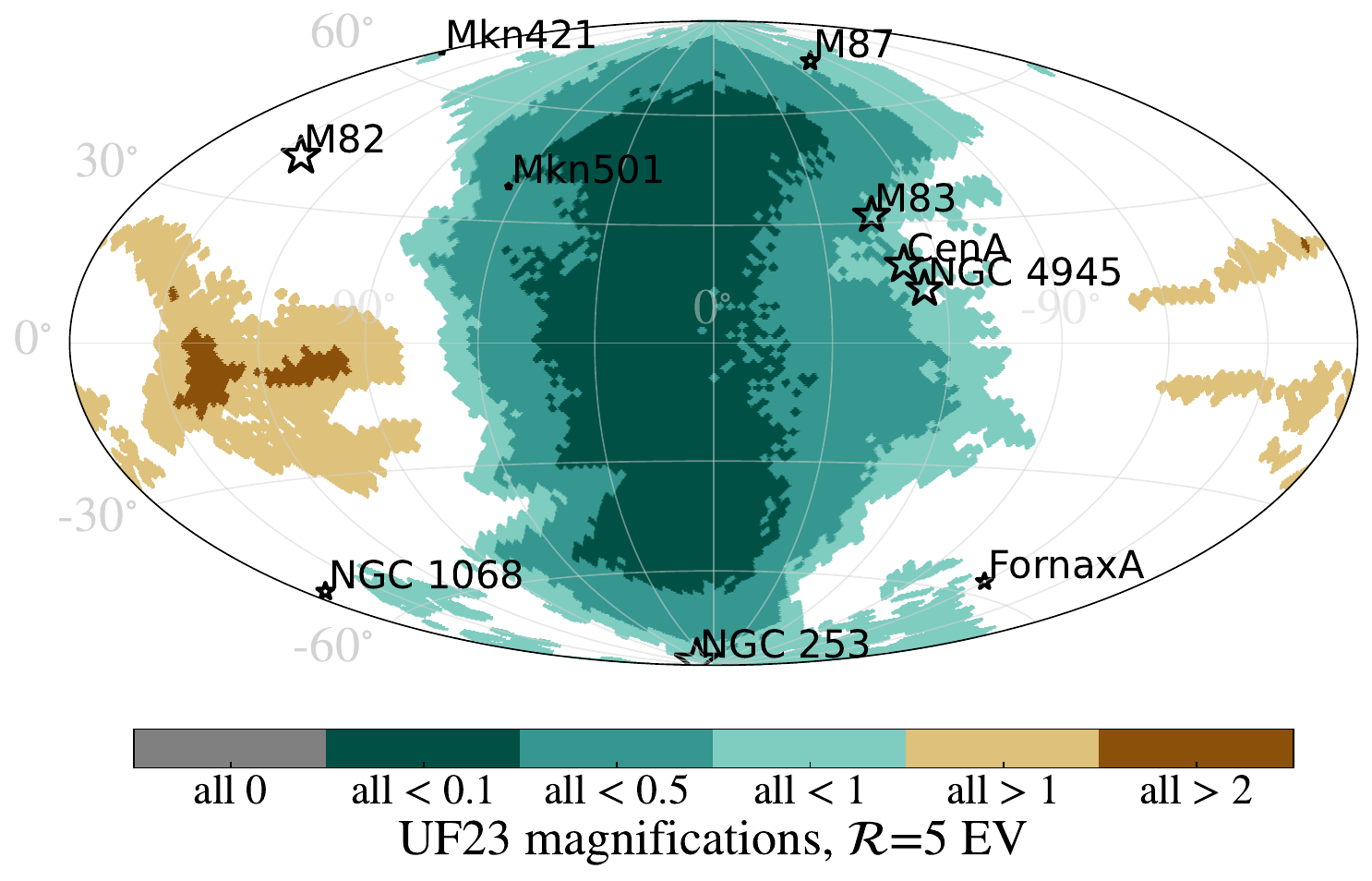}}\hfill
\subfloat[Suppression of a dipole by different GMF models, for an $8.3\%$ perfect extragalactic dipole in the upper plot pointing to the LSS dipole direction (grey square in Fig.~\ref{fig:illumination}), and for the full LSS illumination (Fig.~\ref{fig:illumination}) in the lower plot.]{\label{fig:GMF_sup}\includegraphics[width=0.48\textwidth]{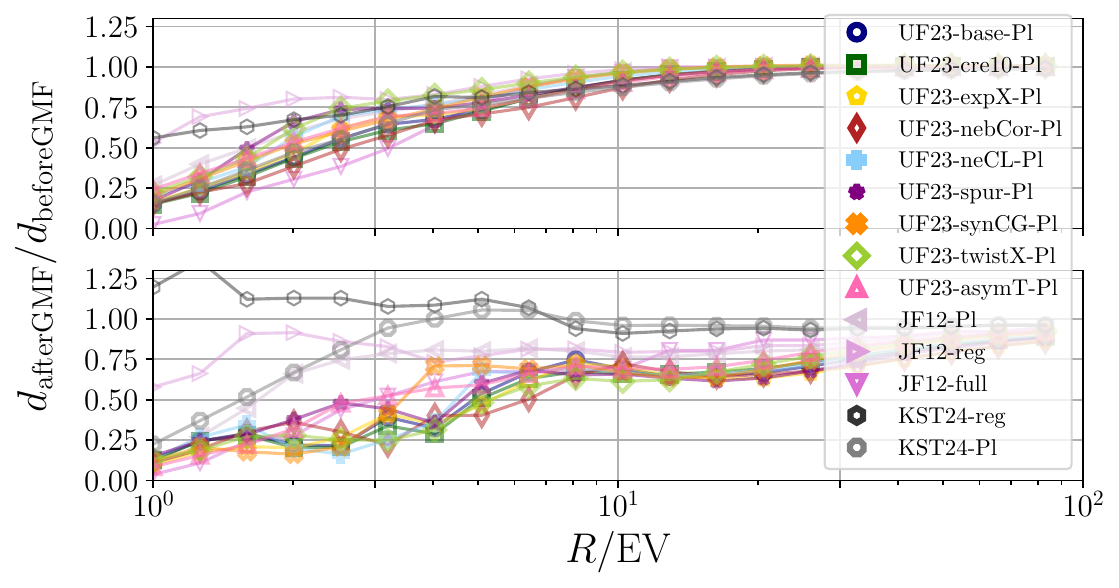}}
\hfill
\subfloat[Deflection directions for different GMF models at $R=20\,\mathrm{EV}$. The CR arrival direction is shown as a filled circle, the origin before different GMF models with different markers, the line indicates smaller rigidities. Combined from~\cite{unger_galactic_2025} and~\cite{korochkin_uhecr_2025}.]{\label{fig:GMF_direc_defl}\includegraphics[width=0.48\textwidth]{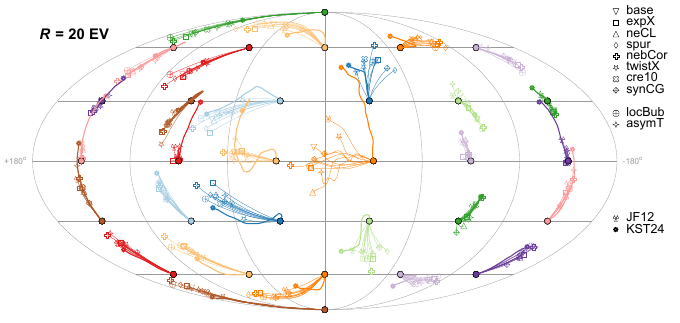}}\hfill
\caption{Impact of the GMF on UHECRs, for details see text.}
\label{fig:GMF}
\end{figure}

The Galactic magnetic field can lead to substantial deflections of UHECRs that have to be accounted for in order to draw conclusions about their sources. The typical deflection of a UHECR with rigidity $R:=E/(e\,Z)=10\,\mathrm{EV}$ is around $30^\circ$ as shown in Fig.~\ref{fig:GMF_median_defl}. Another important effect of the GMF that has to be considered is the (de-)magnification of the flux from different directions~\cite{farrar_deflections_2019, eichmann_galactic_2020, bister_large-scale_2024}: CRs from some directions can reach Earth easily, while CRs e.g. from behind the Galactic center are deflected strongly and simply never reach Earth. This effect is demonstrated in Fig.~\ref{fig:GMF_demag} (for the \texttt{UF23} GMF model suite, see below). It is visible that at $R=5\,\mathrm{EV}$ large parts of the flux from behind the Galactic center and up to the Galactic north and south are strongly demagnified and hence almost invisible to us (see also~\cite{bister_large-scale_2024} for examples with other rigidities). It is important to note that Liouville's theorem~\cite{liouville_nouvelle_1844} is not violated by the flux (de-)magnification. 
An example where demagnification is important is the prediction of large-scale UHECR flux features such as the dipole~\cite{bister_large-scale_2024} as discussed also below in sec.~\ref{sec:LS}. In Fig.~\ref{fig:GMF_sup}, the influence of different GMF models (that will be described below) on an extragalactic flux distribution is shown as a function of rigidity. In the upper plot, that extragalactic flux is solely dipolar and points in the direction of the local extragalactic matter dipole component (shown as a grey square in Fig.~\ref{fig:illumination}, see~\cite{bister_constraints_2024} for details). In general, the GMF dissolves the anisotropy and with that the dipole in the extragalactic flux - an effect that becomes stronger for smaller rigidities where deflections are larger. The GMF influence on the dipolar flux component becomes more complicated when considering also the fine structure in the extragalactic flux (expected from the local matter distribution) instead of just its dipole component as shown in the lower plot. In that case, it can easily happen that the dipole is dissolved more by the GMF at rigidities $R\approx5\,\mathrm{EV}$ than at $R\approx2\,\mathrm{EV}$ (which is the relevant range for UHECRs as described above), or that the dipole amplitude is even intensified by the GMF through magnification (such as for the \texttt{KST24} model).

There are multiple tracers for the magnetic field strength such as Faraday rotation measures or polarized intensity of synchrotron radiation that can be used for modeling the GMF, for a review see e.g.~\cite{jaffe_practical_2019}. The main challenge is that the tracers are only available integrated along the line of sight, which makes the large-scale field structure especially complicated to infer due to Earth's position within the Milky Way disk.
In the last decade, the state of the art GMF model was the \texttt{JF12} model~\cite{jansson_galactic_2012, jansson_new_2012} which consists of parametric model parts for the disk field, as well as the toroidal and poloidal fields. It can be divided into a regular field part responsible for coherent deflections and a random field part. Both are of approximately equal strength. The random field part was updated when new data from the Planck satellite became available~\cite{planck_collaboration_r_adam_et_al_planck_2016}.

In the last two years, great efforts have been made to improve on different aspects of GMF modeling, leading to the emergence of quite some new GMF models. Note however that to this point, only the regular part of the GMF has been studied in these updates. Thus, for now all results shown in this review are produced using the \texttt{JF12} Planck-tuned random field~\cite{planck_collaboration_r_adam_et_al_planck_2016} (abbreviated as "Pl" in all figures).

The \texttt{UF23} suite of GMF models~\cite{unger_coherent_2024} provides for the first time an uncertainty estimate of the GMF model by exploring combinations of different parameterizations, data sets used for the tracers, and auxiliary models such as the model for the distribution of thermal electrons. This has lead to a suite of 8 models which are reasonably representative of the expected uncertainty. Additionally, newer tracer data has been taken into account which is not in agreement with the \texttt{JF12} model anymore. 
The \texttt{KST24} model~\cite{korochkin_coherent_2025} considers for the first time the contribution of the \textit{local bubble} (a magnetized region of $\mathcal{O}(200\,\mathrm{kpc})$ in radius in which the Sun resides~\cite{pelgrims_modeling_2020}) to the tracers during modeling. Additionally, they included the \textit{Fan region} (a region of strong radio emission near the Galactic plane) as a regular Galactic-scale feature instead of masking it out as local foreground - which leads to very strong inferred field strengths in the Perseus spiral arm and is still under debate~\cite{unger_galactic_2025}.
After the release of the \texttt{KST} model, another (preliminary) version within the \texttt{UF23} model suite was produced that also takes into account the effect of the local bubble, called the \texttt{UF23-locBub} model~\cite{unger_galactic_2025}. It results in a slightly better fit quality than the previous 8 \texttt{UF23} models. Note that even when including a more realistic shape for the bubble in variations of the \texttt{UF23} model, the predicted UHECR deflections stay within the ballpark of the other \texttt{UF23} models~\cite{Pelgrims_ICRC_2025}.
After discrepancies in the demagnification of the \texttt{UF23} model suite compared to the predecessor \texttt{JF12} model were found in~\cite{bister_large-scale_2024} (significantly influencing the predicted large-scale UHECR anisotropy, see sec.~\ref{sec:LS}) another model variant of the \texttt{UF23} suite was introduced, the \texttt{UF23-asymT} model~\cite{unger_galactic_2025}. For that model, the \texttt{UF23} fit was forced into a local minimum where the radial extent of the toroidal halo fields is different in the north and south, just as is the case for \texttt{JF12} (and \texttt{KST24}). The \texttt{UF23-asymT} model reaches a $\chi^2$ within the ballpark of the other \texttt{UF23} models, implying that an asymmetric halo cannot be excluded from current rotational measure data.

The deflection of a UHECR with $R=20\,\mathrm{EV}$ within all aforementioned GMF models is depicted in Fig.~\ref{fig:GMF_direc_defl}. In general, a relatively good level of agreement between the models is visible which is promising for UHECR source studies. 
The \texttt{JF12} model generally predicts deflections within the range of the \texttt{UF23} models. The deflections by the \texttt{KST24} model are often in that region as well due to the shared antisymmetric halo, but are generally larger (as expected from Fig.~\ref{fig:GMF_median_defl}). This is especially true in the outer Galaxy due to the strong model field strength in the Fan region.

\subsection{Constraints from large-scale anisotropies} \label{sec:LS}
\begin{figure}[ht]
\centering
\includegraphics[width=0.41\textwidth]{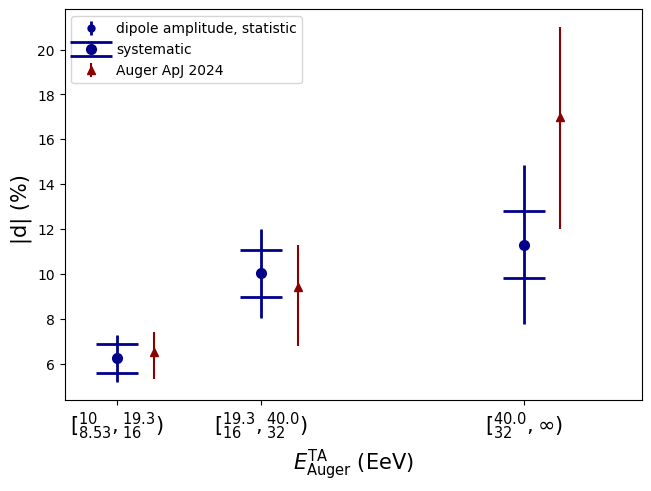}
\includegraphics[width=0.5\textwidth,page=4]{img/aps.pdf}%
\caption{\textit{Left:} amplitude of the dipole moment as measured by Auger only~\cite{abdul_halim_large-scale_2024} (red markers) and by the combined working group of Auger and TA~\cite{rubtsov_update_2025} (blue markers). \textit{Right:} angular power spectrum from the combined Auger+TA data set for the integrated bin corresponding to $E>8.53\,\mathrm{EeV}$ for Auger and $E>10\,\mathrm{EeV}$ for TA. The hatched region (red dashed line) corresponds to the gray solid region (red solid line) but neglecting the systematic uncertainty from the energy scale calibration. Only the dipole deviates from isotropic expectations. Both figures from~\cite{rubtsov_update_2025}.}
\label{fig:dipole_AugerTA}
\end{figure}

\begin{figure}
    \centering
    \subfloat[Dipole direction in Galactic coordinates as measured by Auger only~\cite{abdul_halim_large-scale_2024} (thin ellipses with star centers), and by the combined working group of Auger and TA~\cite{rubtsov_update_2025} (thick ellipses with circle centers) for different energy bins and thresholds.]{\label{fig:dipole_direc}\includegraphics[width=0.54\textwidth]{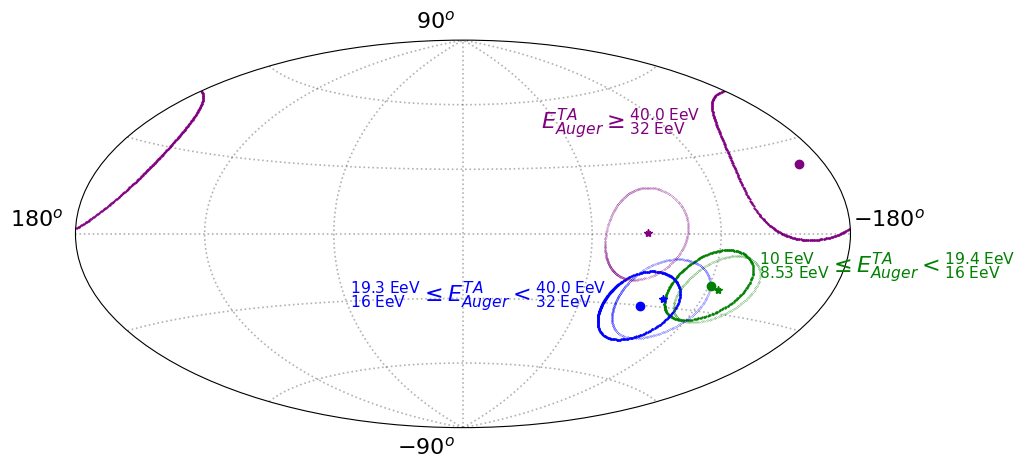}}\hfill
    \subfloat["Illumination": predicted relative UHECR flux at the edge of the Galaxy from the LSS in a mixed-composition hard-injection scenario (see also Fig.~\ref{fig:dipole_comp}), annotated from~\cite{bister_constraints_2024}. The direction of the dipole component is indicated by the grey square.]{\label{fig:illumination}\includegraphics[width=0.44\textwidth]{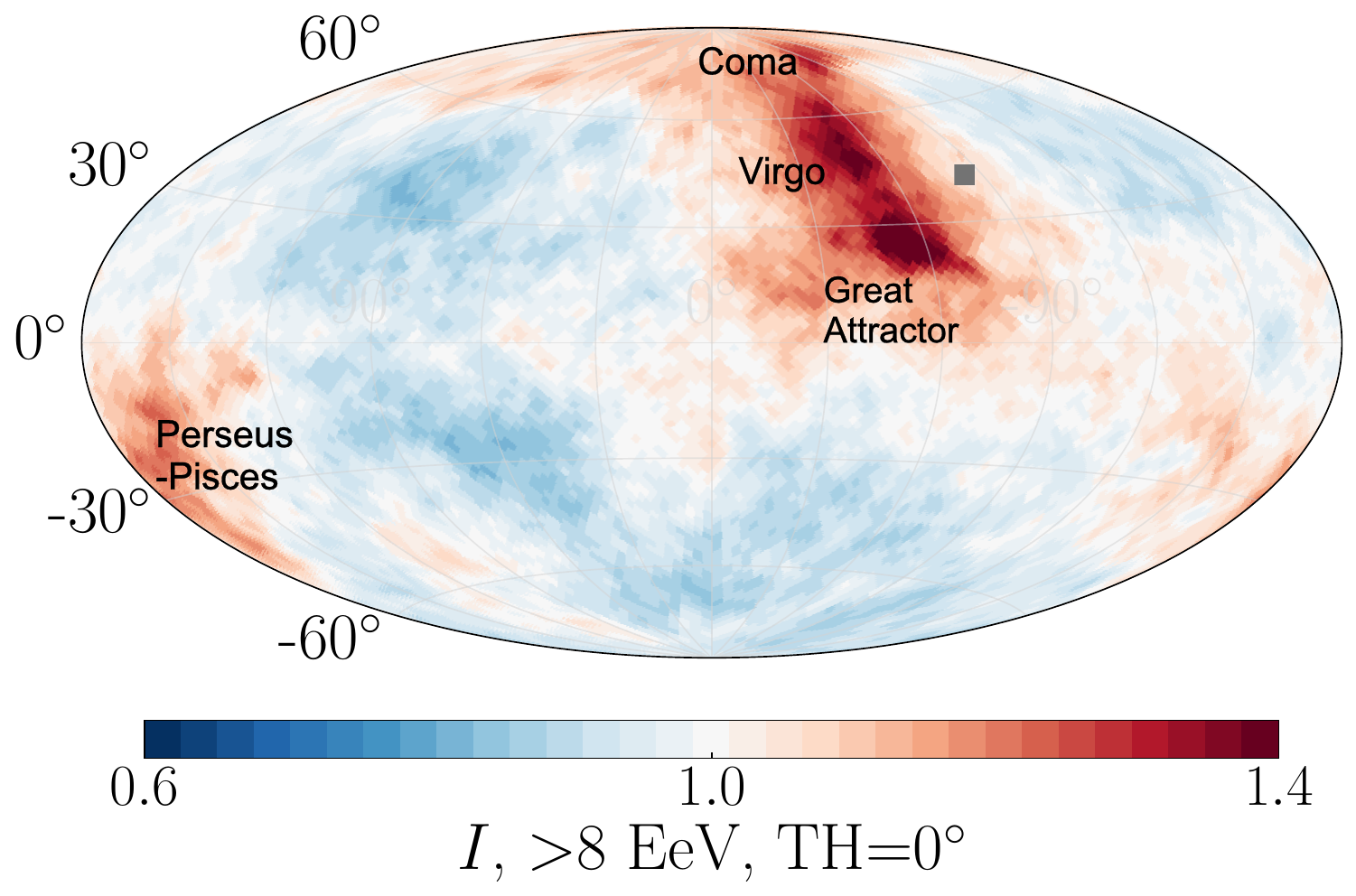}}
    \caption{Measured dipole directions (\textit{left}) and flux predicted for UHECR sources following the LSS (\textit{right}).}
\end{figure}

The only significant anisotropy in the UHECR arrival directions is a large-scale dipole above 8\,EeV~\cite{the_pierre_auger_collaboration_a_aab_et_al_observation_2017}. It has been measured with a current significance of $6.8\sigma$~\cite{abdul_halim_large-scale_2024} as a right-ascension modulation at an energy $E>8\,\mathrm{EeV}$ by Auger.
Using the full-sky data set by Auger+TA~\cite{rubtsov_update_2025}, its amplitude can be constrained without assumptions about the higher moments, to $6.5\%$~\cite{l_caccianiga_on_behalf_of_the_pierre_auger_collaboration_update_2024}. TA data alone is compatible with both isotropy and a dipolar distribution, with a slight preference for the latter~\cite{abbasi_search_2020}.
The dipole amplitude increases with the energy as shown in Fig.~\ref{fig:dipole_AugerTA} (\textit{left}), and the amplitudes are compatible between the Auger-only and Auger+TA measurements. Note that the dipole moment is the only significant one, and that all higher multipole moments are compatible with isotropy as shown in Fig.~\ref{fig:dipole_AugerTA} (\textit{right}).
The dipole direction is shown in Fig.~\ref{fig:dipole_direc}. It points $\sim114^\circ$ away from the Galactic center~\cite{l_caccianiga_on_behalf_of_the_pierre_auger_collaboration_update_2024}, indicating an extragalactic origin of UHECRs at these energies. The dipole direction measured by Auger-only is compatible with the full-sky Auger+TA one apart from the highest energy bin where the angular difference is $\vartheta=81^\circ$. This is due to intermediate-scale flux excesses visible in TA data~\cite{collaboration_indications_2014} (mainly the "TA hotspot" which will be discussed further in sec.~\ref{sec:IS}) driving the direction of the Auger+TA dipole more towards the TA field of view (FOV).

\begin{figure}[ht]
\centering
\includegraphics[width=0.49\textwidth]{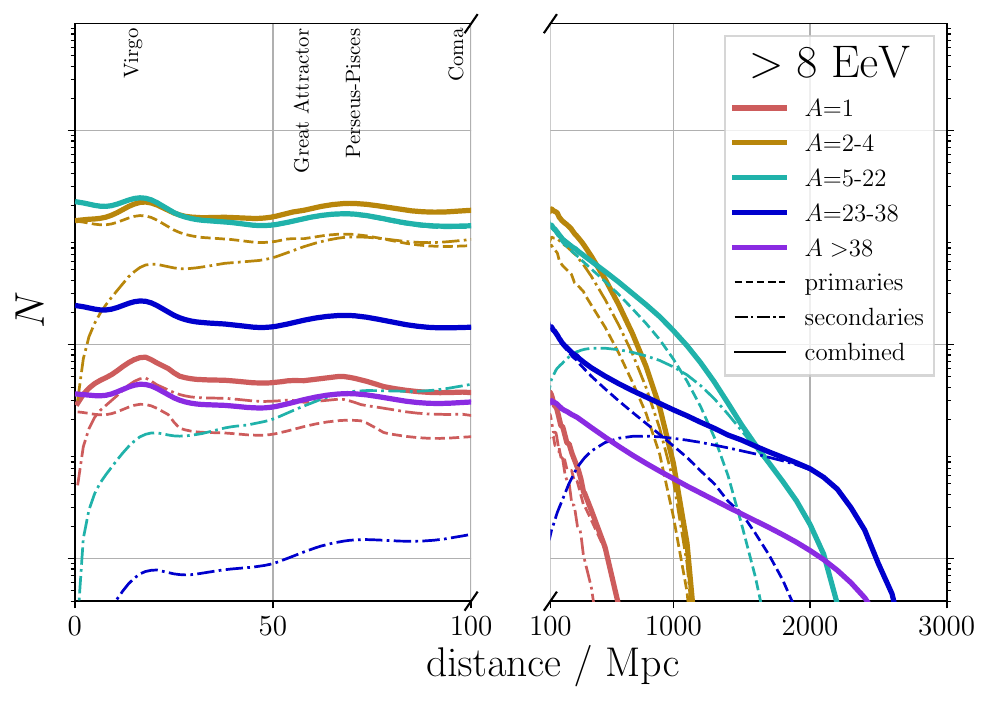}
\includegraphics[width=0.49\textwidth]{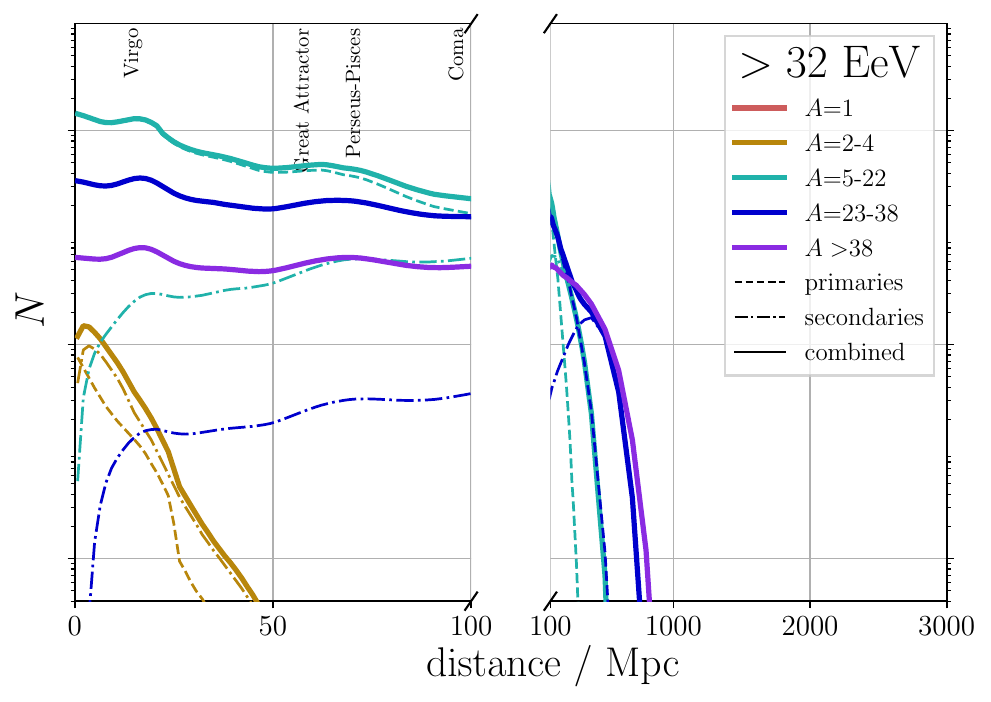}
\caption{Contributions to the flux of different element groups and different distances for two energy thresholds, \textit{left} for $E>8\,\mathrm{EeV}$ and \textit{right} for $E>32\,\mathrm{EeV}$ predicted by the LSS-based model from~\cite{bister_constraints_2024}. The distances of dominant galaxy clusters are indicated at the top. Note that the plots have been split into the nearby (more anisotropic) part $<100\,\mathrm{Mpc}$, and the further away more isotropic part $>100\,\mathrm{Mpc}$, both displayed on linear scales.}
\label{fig:dipole_comp}
\end{figure}

Explaining the dipole has been the aim of several publications in the last decade. Different theories have emerged:
\begin{itemize}
    \vspace{-0.1cm}\item \textbf{A few sources dominate the UHECR flux \boldmath{$>8\,\mathrm{EeV}$} and generate the dipole}. That scenario is for example explored in~\cite{eichmann_ultra-high-energy_2018, eichmann_explaining_2022} where a model based on a few nearby radio galaxies in combination with a diffuse background from farther-away unresolved radio galaxies is used to explain spectrum, composition, and large-scale anisotropies. In most cases discussed in~\cite{eichmann_explaining_2022}, the EGMF has to be rather strong with around $1\,\mathrm{nG}$ for $1\,\mathrm{Mpc}$ coherence length in order not to produce too strong anisotropies, and the flux is dominated by Virgo A (situated in the Virgo cluster) and Fornax A (see also~\cite{seo_energy_2025}). With only few dominant sources, reproducing the subdominant quadrupole, the dipole direction energy evolution, and the spectrum features that are visible over the whole sky~\cite{collaboration_energy_2025, kim_energy_2025} requires finetuning. 
    \vspace{-0.3cm}\item \textbf{An even more extreme case would be that only one source is responsible for the dipole}, a scenario that was explored already long before high-statistics UHECR measurements and models of the GMF were available, see e.g.~\cite{berezinskii_predicted_1990, isola_centaurus_2001}. It is an especially compelling hypothesis because the dominance of a single source would naturally explain the observed similarity of the sources described in sec.~\ref{sec:CF} as well as the directional uniformity of the spectral features.
    Using newest measurements of the spectrum and composition, in~\cite{mollerach_case_2024} a model with the nearest radio galaxy Cen A (at around $3.8\,\mathrm{Mpc}$ distance) supplying the flux above the ankle is tuned to the data. Qualitatively, the model reproduces the large-scale anisotropy for a very strong EGMF field strength of around $(20-50)\,\mathrm{nG}$ that has to be present between Cen A and Earth. If such models can also quantitatively reproduce the arrival directions over the whole energy range will be very interesting to see in the future~\cite{Bister_CenA_2025}.
    \vspace{-0.3cm}\item \textbf{UHECR sources are numerous and at least roughly follow the large-scale structure (LSS)} as explored in~\cite{waxman_signature_1997, cuoco_footprint_2006, harari_anisotropies_2015, tinyakov_full_2015, the_pierre_auger_collaboration_a_aab_et_al_observation_2017, globus_extragalactic_2017, di_matteo_how_2018, the_pierre_auger_collaboration_a_aab_et_al_large-scale_2018, globus_cosmic_2019, ding_imprint_2021, allard_what_2022, bister_constraints_2024, bister_large-scale_2024, abdul_halim_large-scale_2024}. These sources could be compact objects tracing the matter distribution, transient events occurring more often in matter-dense regions, or the acceleration of UHECRs could happen in accretion shocks present around galaxy clusters and filaments~\cite{simeon_hierarchical_2025}.
    The case where UHECR sources are numerous is favored by the fact that the UHECR spectrum is smooth without any bumps hinting at contributions by individual sources.
    In this scenario, the dipole is generated by sources following the anisotropic local matter distribution dominated by galaxy clusters, most importantly Virgo, Great Attractor, and Coma (all in the Galactic north), as well as Perseus-Pisces. This is visible in Fig.~\ref{fig:illumination} showing the flux at the edge of the Galaxy expected from such a model. Note that in such models, the dipole is mostly generated by (sources in) the galaxy clusters in the Galactic north, whose flux is then coherently deflected southwards to reproduce the observed dipole direction (see Fig.~\ref{fig:dipole_direc}). The absence of a flux excess in the Virgo direction that is sometimes regarded as peculiar (e.g.~\cite{fargion_uhecr_2024}) is hence naturally explained by GMF deflections.
    It was demonstrated in~\cite{bister_constraints_2024} that the (dark) matter distribution can be used as a bias-free estimator of the UHECR source distribution, meaning that neither an increased amount of sources in overdense nor in underdense regions is preferred (see also~\cite{waxman_signature_1997}). Especially if Virgo as an overdense cluster region is not emitting UHECRs (e.g. due to magnetic confinement~\cite{condorelli_impact_2023}), the dipole cannot be well reproduced with that model. 
    As the case of multiple sources following the LSS is a natural assumption that is frequently discussed in the literature, and predictions about anisotropies and constraints on quantities such as source density and distribution have been drawn using the newest models of the GMF, the following part of this subsection will concentrate on that scenario.
\end{itemize}
For the large-scale anisotropies to be explainable by sources following the LSS, the composition has to be mixed because a protonic composition is not in agreement with the measured dipole direction~\cite{ding_imprint_2021} and also disfavored by too-large quadrupole moments~\cite{di_matteo_how_2018}. In~\cite{bister_constraints_2024}, a model where the source distribution is assumed to follow the (dark) matter distribution as provided by CosmicFlows2~\cite{hoffman_quasi-linear_2018} is fit to the spectrum, composition, and dipole moments $>8\,\mathrm{EeV}$. The model source emission parameters are in agreement with the findings described in sec.~\ref{sec:CF}. 
The resulting contribution of different element groups from different distances is shown in Fig.~\ref{fig:dipole_comp}. Above 8\,EeV, the flux is dominated by $\sim53\%$ Helium and $\sim41\%$ from the CNO group, while at $>32\,\mathrm{EeV}$, $\sim58\%$ is CNO, $\sim29\%$ Si-like and $\sim12\%$ Fe-like. The effect of different horizons of different elements is visible in Fig.~\ref{fig:dipole_comp}: at these energies, lighter elements like Helium have much smaller interaction lengths than heavier elements, leading to a mass ordering in the distance from which different elements contribute~\cite{batista_quest_2025}. Note that that also implies a significantly larger dipole moment of lighter elements simply from propagation because the anisotropy is generated by the first $\lesssim200\,\mathrm{Mpc}$ after which the universe becomes more and more homogeneous on large scales. In this model, the dipole stems mostly from primaries, moving from a dominant Helium contribution at 8\,EeV to mostly CNO at 32\,EeV. Secondaries, decay products from heavier primaries, typically come from further away as visible in Fig.~\ref{fig:dipole_comp} and are hence more isotropically distributed.

\begin{figure}[ht]
\centering
\includegraphics[width=0.49\textwidth]{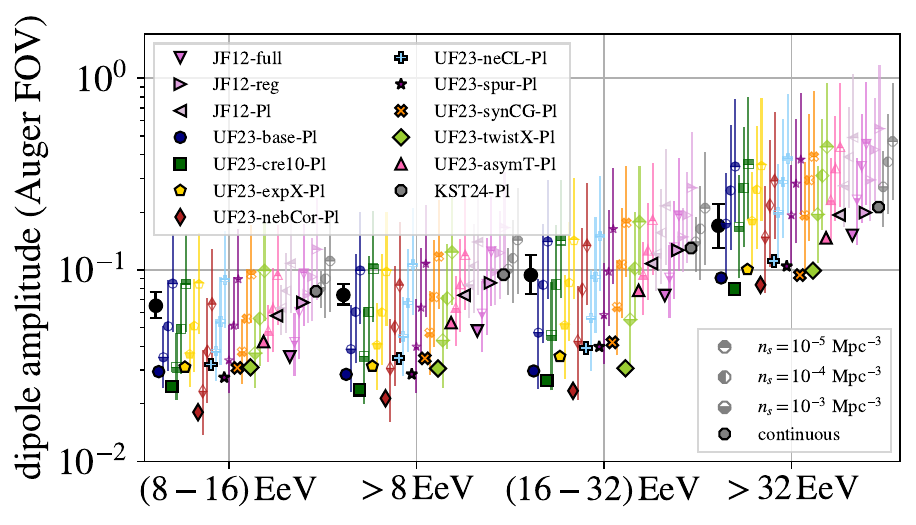}
\includegraphics[width=0.49\textwidth]{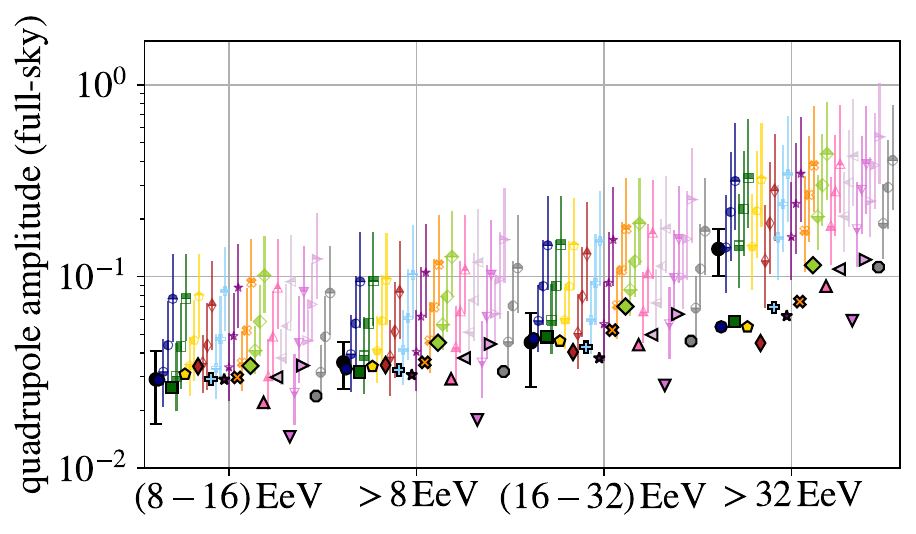}
\caption{Dipole (\textit{left}) and quadrupole (\textit{right}) moments predicted by the LSS model based on CosmicFlows 2~\cite{bister_constraints_2024} using different GMF models, modified from~\cite{bister_large-scale_2024} for different source number densities.
Black circle markers indicate the measured values from~\cite{rubtsov_update_2025} (\textit{right}) and from~\cite{abdul_halim_large-scale_2024} (\textit{left}).}
\label{fig:dipole_predict}
\end{figure}

Another important consequence of this model is that the observed rise of the dipole amplitude with the energy is naturally explained by the shrinking propagation horizon with the energy: only around 25\% of the flux comes from within the more anisotropic $100\,\mathrm{Mpc}$ above 8\,EeV, which increases to 65\% above 32\,EeV as visible in Fig.~\ref{fig:dipole_comp}. As explained in sec.~\ref{sec:CF}, the rigidity stays almost constant with the energy as the mass composition gets heavier, so that the suppression of the dipole for lower rigidities by the GMF as shown in Fig.~\ref{fig:GMF_sup} is only a subdominant effect in this model. 
The evolution of the dipole and quadrupole moments predicted by the LSS-based model from~\cite{bister_constraints_2024} for different GMF models was explored in~\cite{bister_large-scale_2024} and can be seen in Fig.~\ref{fig:dipole_predict} (\textit{upper row}). It is visible that for the continuous model (in the limit of infinite source number density), the dipole amplitude is consistently smaller for the \texttt{UF23} models compared to the predecessor \texttt{JF12} model. This is because the Virgo/Great Attractor overdensity region is demagnified (Fig.~\ref{fig:GMF_demag}) by all \texttt{UF23} models compared to \texttt{JF12}~\cite{allard_what_2024, bister_large-scale_2024}, as visible in Fig.~\ref{fig:GMF_sup}. The underlying reason for this is the symmetric toroidal halo field of the \texttt{UF23} model variants~\cite{unger_galactic_2025} - when using the \texttt{UF23-asymT} model with forced asymmetric halo, the values are indeed more similar to the ones predicted by \texttt{JF12} as visible in Fig.~\ref{fig:dipole_predict}. The \texttt{KST24} model also has an asymptotic halo and magnifies the Virgo/Great Attractor region (see Fig.~\ref{fig:GMF_sup}), leading to a larger dipole component.

When instead of the continuous model, the more appropriate case of a finite source number density is investigated, the dipole and quadrupole moments are influenced by \textit{cosmic variance}: they vary depending on the exact source locations, which is indicated by the $1\sigma$ error bars in Fig.~\ref{fig:dipole_predict}. The smaller the source density, the larger the contribution by less and less individual local sources becomes, driving up the anisotropy. For a source density of around $n_s=\mathcal{O}(10^{-4}\,\mathrm{Mpc}^{-3})$, the best agreement with dipole and quadrupole moments is reached for the \texttt{UF23} models, as larger densities undershoot the dipole and smaller ones overshoot both dipole and quadrupole. For the older \texttt{JF12} model, the \texttt{KST24} model, and the \texttt{UF23-asymT} model, only larger source densities $n_s\gtrsim10^{-3.5}\,\mathrm{Mpc}^{-3}$ are compatible with both quadrupole and dipole amplitudes. This indicates how sensitive the inferred source density is to properties of the GMF model - and that the number has to be treated cautiously, see also the discussion below. Note that these results based on the CosmicFlows~2 LSS model are compatible with studies based instead on the 2MRS galaxy catalog~\cite{allard_what_2022, abdul_halim_large-scale_2024}.
The obtained limits on the source density allow for conclusion on the sources of UHECRs. Sparse source types like blazars ($n_s\sim10^{-7}\,\mathrm{Mpc}^{-3}$~\cite{ajello_cosmic_2013}), high-luminosity AGNs ($n_s\sim10^{-6}\,\mathrm{Mpc}^{-3}$~\cite{gruppioni_herschel_2013}), and even starburst galaxies ($n_s\sim10^{-5}\,\mathrm{Mpc}^{-3}$~\cite{gruppioni_herschel_2013}) may be too sparse to be the sole sources of UHECRs due to too strong anisotropies if no severe smoothing by the EGMF is considered~\cite{bister_constraints_2024, rossoni_investigating_2024}.
More common source types like starforming galaxies, low-luminosity AGNs, or transients with high rates like non-jetted TDEs~\cite{farrar_giant_2009} or BNS merger~\cite{farrar_binary_2025} are instead preferred.

\begin{figure}[ht]
\subfloat[energy $>8\,\mathrm{EeV}$]{\includegraphics[trim={11.0cm 0 0 0}, clip, height=4cm]{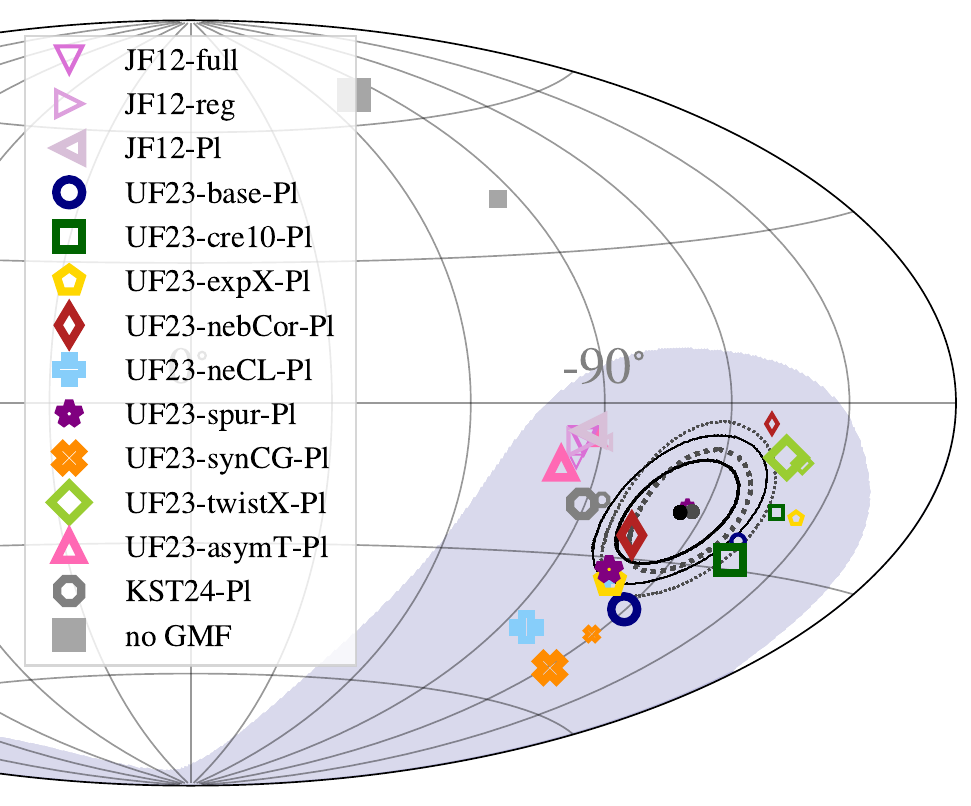}}
\subfloat[$(8-16)\,\mathrm{EeV}$]{\includegraphics[trim={15.6cm 0 0 0}, clip, height=4cm]{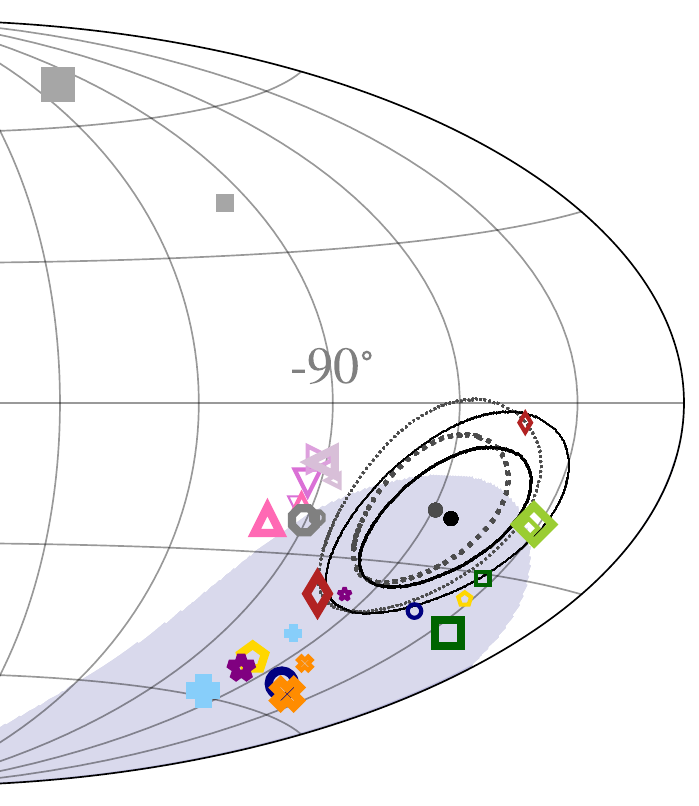}}
\subfloat[$(16-32)\,\mathrm{EeV}$]{\includegraphics[trim={15.6cm 0 0 0}, clip, height=4cm]{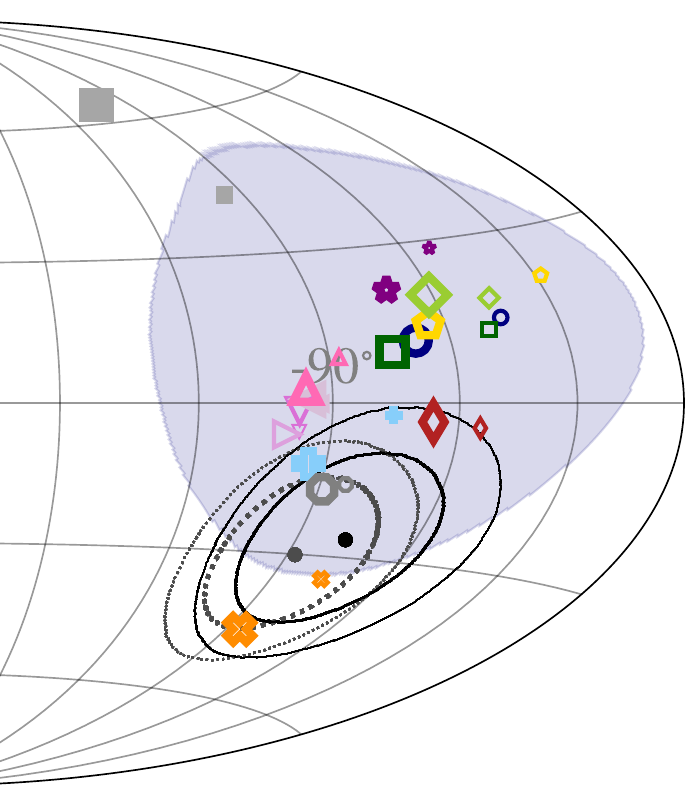}}
\subfloat[$>32\,\mathrm{EeV}$]{\includegraphics[trim={15.6cm 0 0 0}, clip, height=4cm]{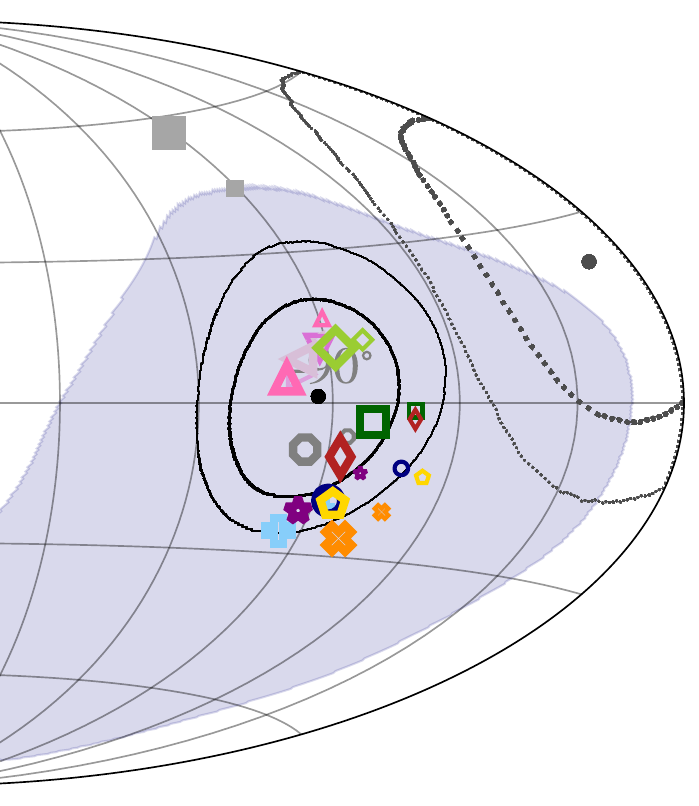}}
\caption{Predicted and measured dipole directions in Galactic coordinates, modified from~\cite{bister_large-scale_2024}:
Colored markers indicate the dipole directions for different GMF models where larger markers show the Auger FOV and smaller markers the full-sky one. The blue region shows the $1\sigma$ uncertainty due to cosmic variance in the source positions, for the \texttt{UF23-base} model with $n_s=10^{-3}\,\mathrm{Mpc}^{-3}$ and using the Auger FOV. The black contours represent the $1\sigma$ and $2\sigma$ uncertainty domains of the measured dipole in the Auger FOV~\cite{abdul_halim_large-scale_2024}, and the gray dotted ones for the whole sky as measured by Auger+TA~\cite{rubtsov_update_2025}.}
\label{fig:dipole_predict_direc}
\end{figure}

The predicted dipole directions of the CosmicFlows-based model are shown in Fig.~\ref{fig:dipole_predict_direc}. Note that the dipole directions predicted using the 2MRS catalog~\cite{abdul_halim_large-scale_2024, allard_what_2022} agree with the CosmicFlows predictions shown in Fig.~\ref{fig:dipole_predict_direc} within $\mathcal{O}(10^\circ)$ for the \texttt{JF12} model. Predictions for other GMF models based on the 2MRS catalog are not yet available.
The dipole directions predicted using the full sky and using only the limited Auger FOV (both depicted in Fig.~\ref{fig:dipole_predict_direc}) can differ by up to $\vartheta=30^\circ$ at lower energies, depending on the GMF model. For the highest energy bin $E>32\,\mathrm{EeV}$, the difference is $\vartheta\lesssim10^\circ$ for all models.
All GMF model predictions are roughly compatible with the measured dipole directions by Auger (even though no model fits perfectly for all energies). However, none of the models can reproduce the full-sky measured dipole direction by Auger+TA\footnote{Note that the measured $\vartheta=81^\circ$ angular difference between the full-sky and Auger FOV dipole directions in the $E>32\,\mathrm{EeV}$ bin is also not reproduced when considering cosmic variance. On average, it is$\sim20^\circ$ for $n_s=10^{-3}\,\mathrm{Mpc}^{-3}$ and $\lesssim30^\circ$ for $n_s=10^{-4}\,\mathrm{Mpc}^{-3}$, independently of the GMF model.} in the highest energy bin $>32\,\mathrm{EeV}$, that differs substantially from the Auger FOV one. This is because the intermediate-scale anisotropies measured by TA at energies $\gtrsim40\,\mathrm{EeV}$ (see sec.~\ref{sec:IS}) pulling the dipole direction closer to the TA FOV are not reproduced by the LSS with any GMF model. Thus, if these intermediate scale anisotropies turn out to be statistically significant, they are more likely to originate from local or transient sources instead of the galaxy clusters represented in the LSS, see also~\cite{he_monte_2016}.

The difference between predictions using different random field coherence lengths is $\mathcal{O}(15^\circ)$ at lower energies and decreases to $\mathcal{O}(5^\circ)$ for $E>32\,\mathrm{EeV}$ (not shown, see~\cite{bister_large-scale_2024}).
The predicted dipole direction clearly also depends on cosmic variance, and variations increase fast with decreasing source density. In Fig.~\ref{fig:dipole_predict_direc}, the $1\sigma$ uncertainty for the \texttt{UF23-base} model at $n_s=10^{-3}\,\mathrm{Mpc}^{-3}$ is shown exemplarily. Already at that source density, the uncertainty from cosmic variance is larger than the differences between different GMF models.

Next to the uncertainty from cosmic variance which is an inherent uncertainty that is not avoidable until the sources of UHECRs are identified individually, there are evidently other uncertainties involved in models such as the ones described above. The uncertainty on the coherent GMF model is manageable thanks to the recent development of several new models~\cite{bister_large-scale_2024}. Remarkably, even GMF models by different authors agree on both the dipole direction and amplitude predictions within the uncertainties from cosmic variance for source density $\lesssim10^{-2}\,\mathrm{Mpc}^{-3}$. The EGMF, however, could have an unknown effect, both on the inferred source density and also on other model parameters as investigated in~\cite{bister_constraints_2024, mollerach_extragalactic_2020, abdul_halim_impact_2024, allard_what_2022}. But, to estimate that effect reliably, more knowledge about the structure and strength of the EGMF~\cite{durrer_cosmological_2013, hackstein_simulations_2018, vazza_magnetogenesis_2021}, that of nearby galaxy clusters~\cite{condorelli_impact_2023}, and the magnetic field of the local group are necessary. When considering also structured EGMFs instead of simple turbulent approximations, simulations and model fits become extremely computationally expensive. 

Another important uncertainty stems from the source model - while it seems that conclusions reached with models based on LSS approximations from dark-matter distributions such as CosmicFlows 2~\cite{globus_extragalactic_2017, globus_cosmic_2019, ding_imprint_2021, bister_constraints_2024, bister_large-scale_2024} agree in general with the conclusions drawn from models based on galaxy catalogs~\cite{di_matteo_how_2018, abdul_halim_large-scale_2024, allard_what_2022}, it was also shown that the predicted large-scale anisotropies are extremely sensitive to the interplay of GMF model demagnification and source distribution.
Updated predictions using CosmicFlows 4~\cite{valade_identification_2024} as well as explicit galaxy catalogs with newer GMF models are needed for better estimation of the impact of the source distribution. Note that also a homogeneous source distribution can lead to the observed large-scale anisotropies for sufficiently small source densities $n_s\lesssim10^{-4}\,\mathrm{Mpc}^{-3}$~\cite{harari_anisotropies_2015, the_pierre_auger_collaboration_a_aab_et_al_large-scale_2018, lang_ultrahigh-energy_2021, allard_what_2022, bister_constraints_2024}, but for that case the dipole direction is clearly non-informative.

Finally, the mass composition can also have significant impact on the predicted large-scale anisotropies. While variations of the hadronic interaction model or source emission fit parameters seem to have minor impact~\cite{bister_constraints_2024, allard_what_2022}, significant shifts of the \xmax{} scale towards heavier elements could lead to a decreased level of anisotropy~\cite{vicha_heavy-metal_2025}.
The measurement of mass estimators for individual events at the highest energies~\cite{pierre_auger_collaboration_inference_2025} also offers great chances to learn more about UHECR sources and the GMF. E.g. the analysis proposed in~\cite{martins_prospects_2025} where the dipole component is calculated for a heavy and a light event selection in each energy bin could test the mass-dependent effects described above: the anticorrelation between dipole moment and mass expected from propagation, and the rigidity-dependent suppression by the GMF as shown in Fig.~\ref{fig:GMF_sup}. Especially a significant split of the dipole direction between heavy and light selections could be extremely informative about GMF models and possibly even the influence of local sources.

\subsection{Constraints from intermediate-scale anisotropies} \label{sec:IS}
\begin{figure}[th]
\centering
\includegraphics[width=0.5\textwidth]{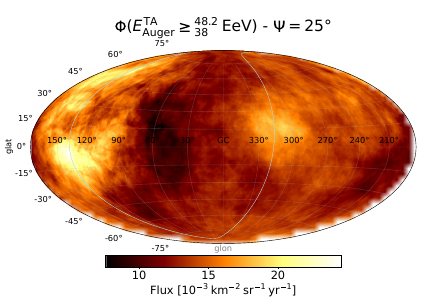}
\includegraphics[width=0.48\textwidth]{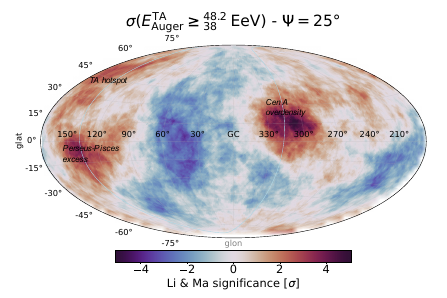}
\caption{\textit{Left:} Measured arrival flux of UHECRs in Galactic coordinates with $E\gtrsim40\,\mathrm{EeV}$ (see title) with a $25^\circ$ tophat blurring. \textit{Right:} Corresponding local LiMa significance. The names often used in the literature for the three prominent overdensities are indicated. Both (modified) from \cite{l_caccianiga_on_behalf_of_the_pierre_auger_collaboration_update_2024}.}
\label{fig:ADs_40_allsky}
\end{figure}

While at energies $\sim8\,\mathrm{EeV}$ the UHECR sky is dominated by the large-scale dipole signal, at higher energies $\gtrsim32\,\mathrm{EeV}$ anisotropies at smaller scales start arising. This is as expected from the shrinking propagation horizon as visible in Fig.~\ref{fig:dipole_comp}, indicating a larger relative flux contribution of local sources. The full-sky flux at energies $\gtrsim40\,\mathrm{EeV}$ as measured by Auger and TA combined is shown in Fig.~\ref{fig:ADs_40_allsky} (\textit{left}). 
The Auger data has been scanned for anisotropies over different angular scales and energy thresholds in every available direction, revealing no significant anisotropies beyond a $p$-value of $2\%$~\cite{g_golup_on_behalf_of_the_pierre_auger_collaboration_update_2024, the_pierre_auger_collaboration_p_abreu_arrival_2022}. The maximum significance is reached at $E>38\,\mathrm{EeV}$ for an angular scale of $27^\circ$. It is close to the direction of the nearby radio galaxy Cen A, which will be discussed further below. The autocorrelation and correlation with the Galactic and supergalactic plane all yield $p$-values $>10\%$. 
In the TA data, a scan has revealed an overdensity at $E>57\,\mathrm{EeV}$ which is often referred to as the "TA hotspot"~\cite{collaboration_indications_2014}. Its significance grows over time and is now at $2.9\sigma$ post-trial on an angular scale of $25^\circ$~\cite{kim_telescope_2025}. Several theories for its origin have been discussed in the literature, mostly nearby candidates close to the hotspot direction~\cite{he_monte_2016} like the starburst galaxy M82, or more complicated proposals like CRs originating in Virgo and then being deflected along filaments~\cite{kim_filaments_2019} (note however that a pure-proton composition was used in~\cite{kim_filaments_2019}).
Another excess at slightly lower energies $\sim25\,\mathrm{EeV}$ was later identified in a manual search at $20^\circ$ angular windows with a local significance of $\sim3.9\sigma$. Often, this overdensity is referred to as the "Perseus-Pisces" (PP) excess because it is close to the supercluster direction (compare to Fig.~\ref{fig:illumination}). Note that the quoted post-trial significance of $3.2\sigma$~\cite{kim_telescope_2025} refers to how often such an excess appears close to PP instead of quantifying the overdensity on a statistical basis accounting for the scan in direction and energy that was indirectly performed. Note also that in the studies described above based on a LSS model including PP as a dominant source in the extragalactic flux, its flux excess is dissolved by the GMF and not observable at Earth~\cite{bister_constraints_2024}.

The combined full-sky significance map from Auger and TA is shown in Fig.~\ref{fig:ADs_40_allsky} (\textit{right}) with the names of the three described intermediate-scale overdensities indicated. Interestingly, both the TA hotspot and the PP-excess are also within the FOV of Auger, albeit at the border and only observed in inclined events. However, due to the larger size of Auger, its exposure in the two excess regions is comparable to the one of TA. In Auger data, no overdensity in neither the hotspot nor the PP direction is observed~\cite{g_golup_on_behalf_of_the_pierre_auger_collaboration_update_2024, the_pierre_auger_collaboration_flux_2024}. 
This also disfavors the claim~\cite{collaboration_observation_2024} that flux excesses in the northern hemisphere only visible for TA could explain the discrepancy in the flux seen by Auger and TA, a hypothesis investigated e.g. in~\cite{globus_can_2017, plotko_differences_2023}. Note that taking into account also possible energy scale mismatches, a mere statistical overfluctuation seen by TA, or an underfluctuation seen by Auger in the excess regions cannot be excluded at this point~\cite{the_pierre_auger_collaboration_flux_2024}.

\paragraph{\textbf{Correlation with Cen A:}} The correlation with the nearby radio galaxy Cen A is currently observed with $4\sigma$ level post-trial significance for $E>38\,\mathrm{EeV}$~\cite{g_golup_on_behalf_of_the_pierre_auger_collaboration_update_2024} at an angular scale of $27^\circ$. Cen A has been suspected as a source of UHECRs for years~\cite{cavallo_sources_1978, romero_centaurus_1996, farrar_deducing_2000, gorbunov_comment_2007}, even long before the overdensity had emerged, due to its proximity and powerful radio jets. Acceleration of UHECRs up to the highest energies is possible for Cen A according to simulations and theoretical modeling, e.g.~\cite{wykes_mass_2013, wang_acceleration_2024}.

In~\cite{the_pierre_auger_collaboration_a_abdul_halim_et_al_constraining_2024}, a model was simultaneously fit to the Auger energy spectrum, mass composition, and arrival directions $>16\,\mathrm{EeV}$, based on the assumption that Cen A (or other galaxy catalogs, see below) is a source of UHECRs and contributes part of the flux above the ankle. Additionally, another part of the model flux originates from homogeneous background sources, as in the models introduced above in sec.~\ref{sec:CF}. By taking into account propagation effects and a rigidity-dependent blurring (but no coherent deflections), the level of anisotropy rises naturally with the energy due to the increased relative contribution of the nearby source Cen A with a best-fit of around $3\%$ at $40\,\mathrm{EeV}$.
The model including Cen A describes mostly the arrival direction distribution significantly better than the model based on only homogeneous sources, especially around $10^{19.3}\,\mathrm{eV}\approx20\,\mathrm{EeV}$ and $\sim10^{19.7}\,\mathrm{eV}\approx50\,\mathrm{EeV}$. These two peaks in the correlation could hint at different mass groups being accelerated to the same rigidity~\cite{lemoine_anisotropy_2009}. In the model discussed in~\cite{the_pierre_auger_collaboration_a_abdul_halim_et_al_constraining_2024}, the dominant mass groups at the respective energies are Helium and CNO with an appropriate factor 2-3 between their charge numbers.

\begin{figure}[ht]
\subfloat[$Z=2$, $\overline{R}\approx26.8\,\mathrm{EV}$]{\includegraphics[trim={15cm 0 0 0}, clip, height=4cm]{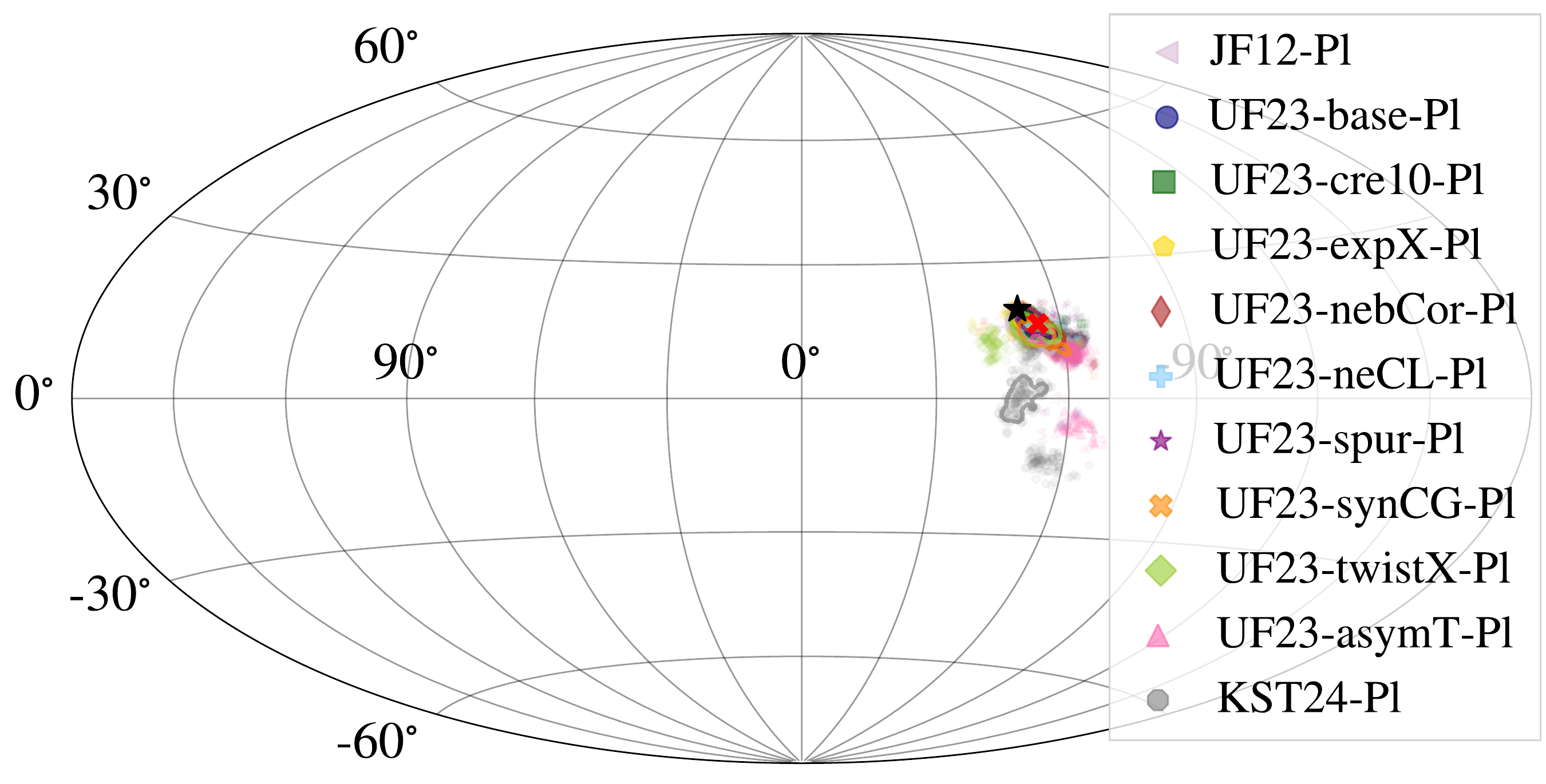}}\vspace{0.1cm}
\subfloat[$Z=6$, $\overline{R}\approx9\,\mathrm{EV}$]{\includegraphics[trim={15cm 0 0 0}, clip, height=4cm]{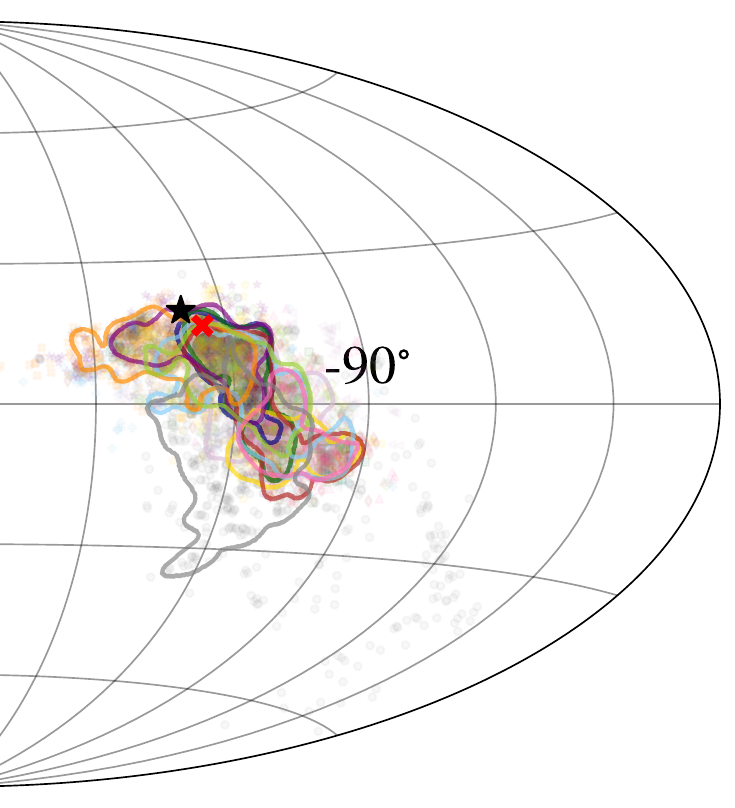}}
\subfloat[$Z=12$, $\overline{R}\approx4.5\,\mathrm{EV}$]{\includegraphics[trim={15cm 0 0 0}, clip, height=4cm]{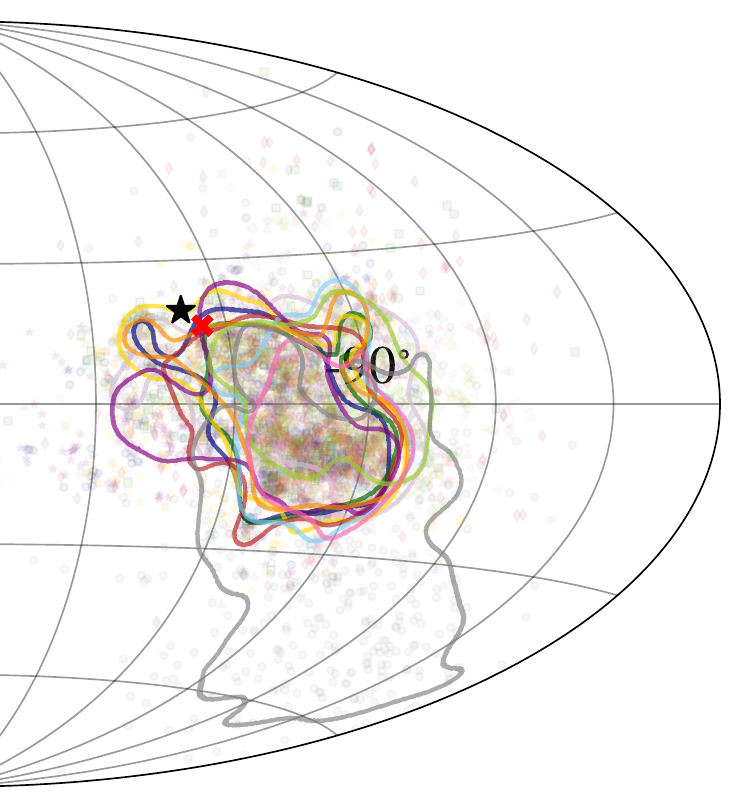}}
\subfloat[$Z=26$, $\overline{R}\approx2\,\mathrm{EV}$]{\includegraphics[trim={15cm 0 0 0}, clip, height=4cm]{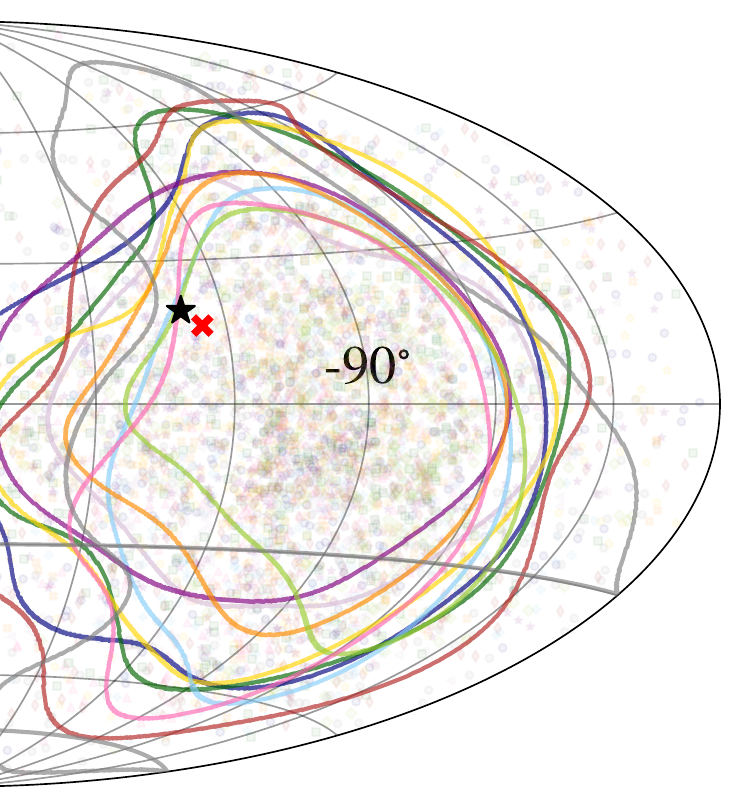}}
\caption{Simulated UHECRs from Cen A (black star) using different GMF models. The energies of the events follow the Auger spectrum $>38\,\mathrm{EeV}$ and the charge number and resulting mean rigidity are given below each plot. The contours indicate the regions containing $90\%$ of all events. The red cross indicates the direction of the maximum found in the Auger overdensity search~\cite{the_pierre_auger_collaboration_p_abreu_arrival_2022}.}
\label{fig:cena_defl}
\end{figure}

In~\cite{the_pierre_auger_collaboration_flux_2024}, it was recently confirmed that the Cen A excess extends to smaller energies of $\sim20\,\mathrm{EeV}$. The position of the excess remains constant with the energy, which could indicate that the CR rigidity remains constant or that no strong coherent deflections occur due to a subdominant regular magnetic field in that direction. Note that the angular window was kept fixed in the analysis~\cite{the_pierre_auger_collaboration_flux_2024}, so no conclusions regarding the GMF can be drawn from that. 
The arrival directions of UHECRs from Cen A that are expected for different GMF models are shown in Fig.~\ref{fig:cena_defl}. It is visible that all GMF models predict that CRs are deflected towards the Galactic plane, which is also the direction where the maximum overdensity occurs, see the cross marker in Fig.~\ref{fig:cena_defl}. Overall, coherent displacement remains rather small, for all GMF models apart from the \texttt{KST24} model. Comparing these simulations visually to Fig.~\ref{fig:ADs_40_allsky}, it does seem reasonable that Cen A could be the origin of the flux excess if at least part of its flux consists of events with $R\gg4\,\mathrm{EV}$, possibly in combination with some additional blurring from a turbulent EGMF that is not considered in the simulations shown in Fig.~\ref{fig:cena_defl}, see also~\cite{dolgikh_cen_2025} and~\cite{Bister_CenA_2025}.

\paragraph{\textbf{Correlation with active galactic nuclei:}}
Not just the nearest radio galaxy Cen A, but also active galactic nuclei (AGNs) in general were long suspected to be sources of UHECRs due to their powerful jets and magnetized lobes, both of which could be good candidates for acceleration to the highest energies~\cite{matthews_particle_2020, rieger_active_2022}. Different theories for acceleration in the core, lobes, and jets of AGNs (including also two-stage and re-acceleration processes) have been investigated, see e.g.~\cite{caprioli_espresso_2015, murase_high-energy_2022, kimura_ultrahigh-energy_2018, matthews_ultrahigh_2019, ehlert_ultra-high-energy_2025}. 
In the case that UHECRs are accelerated along the jet, blazars where the jet points into the direction of Earth are prime candidates for UHECR acceleration~\cite{resconi_connecting_2017, rodrigues_active_2021}. Note that the detection of a correlation of the UHECR arríval directions with AGNs might be hindered if time delays e.g. due to the EGMF become larger than typical AGN duty cycles~\cite{mbarek_revisiting_2025}.

Early Auger data $>60\,\mathrm{EeV}$ seemed to correlate with AGNs on angular scales $\lesssim6^\circ$~\cite{the_pierre_auger_collaboration_j_abraham_correlation_2007}. While that correlation turned out to be a statistical fluctuation, current tests for correlation between the Auger+TA arrival directions and AGNs have reached a significance of $3.3\sigma$. 
Three parameters were optimized, the energy threshold $E>38\,\mathrm{EeV}$, an anisotropic fraction $4.8\%$ and a Fisher search radius around the sources of $15.4^\circ$~\cite{Galvez_ICRC_2025}. Additionally, a model of jetted AGNs ($\gamma$-AGNs) has been tested, where the UHECR luminosity is assumed to be scaled with the $\gamma$-ray flux measured by Fermi. The model reaches a significance of $\sim3.8\sigma$ for a similar angular scale and energy threshold and an anisotropic fraction of $8.8\%$. In the subsequent analysis including an energy-dependent modeling of the catalog contribution in a combined fit to spectrum, composition, and arrival directions measured by Auger, the $\gamma$-AGN model was however disfavored due to the strong dominance of the far-away blazar Markarian 421~\cite{the_pierre_auger_collaboration_a_abdul_halim_et_al_constraining_2024}. In~\cite{de_oliveira_gamma_2025} it was suggested to correct the UHECR flux weighting with the observed $\gamma$-ray flux, as that implicitly assumes that UHECRs experience the same beaming along the jet as $\gamma$-rays - which is unexpected due to magnetic field decollimation. Instead, for example the intrinsic unbeamed $\gamma$-ray flux or the jet power could be used as a proxy. This is also in accordance with the finding from~\cite{eichmann_ultra-high-energy_2018, eichmann_explaining_2022} that nearby radio galaxies could explain the measured Auger data including the dipole when the UHECR flux is more carefully modeled and not simply scaled with the $\gamma$-ray flux (see sec.~\ref{sec:LS}), and the observation that $\gamma$-ray bright sources like blazars alone are too sparse and hence produce too much anisotropy~\cite{partenheimer_ultra-high-energy_2024, bister_constraints_2024}. Instead of considering only the brightest and most powerful AGNs that lead to too large anisotropies, lately also models considering weaker more prevalent classes of AGNs like FR0~\cite{merten_scrutinizing_2021, lundquist_combined_2025} or Seyfert galaxies~\cite{anjos_central_2017} have been investigated, even though it remains unclear if they can produce sufficiently high energies. Those weaker source candidates also have the advantage that their source evolution is weaker~\cite{hasinger_luminosity-dependent_2005} and hence does not lead to the overproduction of secondaries overshooting the spectrum below the ankle as is the case for powerful AGNs~\cite{the_pierre_auger_collaboration_a_abdul_halim_constraining_2023, the_pierre_auger_collaboration_a_abdul_halim_et_al_constraining_2024}, see sec.~\ref{sec:CF}.

\paragraph{\textbf{Correlation with Starburst galaxies:}}
Next to AGNs, also a catalog of starburst galaxies (SBGs) has been tested for correlations with arrival direction data. These are galaxies detectable by high infrared luminosities as they undergo extreme starformation activity, driving powerful magnetized winds. Comparing the Auger+TA arrival direction data to a model where part of the UHECR flux comes from these SBGs, currently a significance of $4.2\sigma$ is reached for an energy threshold $E>38\,\mathrm{EeV}$, an anisotropic fraction $\alpha=10.6\%$ and a search radius of $17.6^\circ$~\cite{Galvez_ICRC_2025}.
Using again the model including propagation and rigidity-dependent blurring for a combined fit of the Auger energy spectrum, composition, and arrival directions, the SBG model is favored 
over a homogeneous model at $4.5\sigma$ significance (including experimental systematic uncertainties)~\cite{the_pierre_auger_collaboration_a_abdul_halim_et_al_constraining_2024}. Especially the excess in the Cen A region is well described by the SBG model. In that model, the flux excess comes from NGC4945 instead of Cen A, an SBG close in direction and at a similar distance $\sim4\,\mathrm{Mpc}$.

Despite its quite large significance, the observed correlation with SBGs is difficult to interpret because only symmetric search radii or a rigidity-dependent blurring was taken into account in the analyses~\cite{Galvez_ICRC_2025, the_pierre_auger_collaboration_a_aab_et_al_indication_2018, the_pierre_auger_collaboration_a_abdul_halim_et_al_constraining_2024}, but no coherent displacements as expected by the regular part of the GMF. In~\cite{allard_what_2024, m_kuznetsov_on_behalf_of_the_pierre_auger_collaboration_possible_2024}, it was shown that it is possible but not straightforward to reproduce the SBG correlation if UHECR sources simply follow the 2MRS catalog and GMF deflections are included. The correlation is however easier to reproduce in models with coherent deflections if UHECR sources are galaxies with high starformation rate~\cite{allard_what_2024}.
In~\cite{higuchi_effects_2023, deval_impact_2025}, it was also shown that it is possible to reproduce the significance, the best-fit search radius, and the anisotropic fraction inferred by Auger in simulations based on the SBG catalog even when including a coherent GMF model. In that case, the true signal fraction from the SBG catalog has to be higher $\sim(10-30)\%$ (depending on the GMF model) than the inferred anisotropic fraction of $\sim10\%$ because heavier particles are deflected too much and do not fall within the search radius anymore~\cite{deval_impact_2025}.

Naturally, the observed correlation between the UHECR data and starburst galaxies has sparked debate if SBGs could actually be the sources of UHECRs~\cite{lunardini_are_2019, vanvliet_extragalactic_2022, oliveira_nearby_2023}. It is still somewhat unclear if and how CRs could be accelerated to ultra-high energies in SBGs, especially because strong radiation losses are expected~\cite{anchordoqui_acceleration_2018, romero_particle_2018, anchordoqui_exploring_2020, peretti_exploring_2021, condorelli_testing_2023}. Alternatively, UHECRs could also be (pre-)accelerated not by the SBGs themselves, but by transient events like supernovae or GRBs which are very numerous in the central region of SBGs~\cite{kachelriess_extragalactic_2022}. 
The hypothesis that transient events occuring proportionally to the SFR in any galaxy are the sources of UHECRs is investigated in~\cite{marafico_closing_2024}. By comparing models with different transient burst rates to the flux excesses seen in the Cen A region and the TA hotspot region, they conclude that long GRBs are favored to explain the measured data. Note however that that conclusion may change once coherent magnetic field deflections and event-by-event data is taken into account~\cite{bister_prospects_2025}.
Another possibility to explain the observed correlation of UHECRs with SBGs without SBGs being the actual sources of UHECRs is explored in~\cite{bell_echoes_2022, taylor_uhecr_2023}. In that model, UHECRs are accelerated by Cen A in an earlier powerful outburst episode and then merely scattered by magnetic fields of the nearby galaxies forming the "Council of Giants" - which includes many of the SBGs used in the correlation analysis described above~\cite{the_pierre_auger_collaboration_a_aab_et_al_indication_2018, the_pierre_auger_collaboration_p_abreu_arrival_2022}.

\paragraph{\textbf{Deflection patterns and multiplets:}}
While many of the described analyses hint at nearby source candidates like Cen A or SBGs being sources of UHECRs, no firm conclusions can be drawn yet.
Even if a correlation significance surpasses $5\sigma$, it could still be that the overdensity is actually caused by other sources. For the Cen A excess for example, other source candidates next to Cen A, such as the nearby starburst galaxy NGC4945 (see above), or further away sources like the Virgo cluster could also explain the observed anisotropies~\cite{ding_imprint_2021, allard_what_2022, allard_what_2024}.

One way to disentangle the different sources in the Cen A excess region would be to search for energy-ordered events (\textit{multiplets}) as expected from magnetic field deflections. In a blind search, no significant multiplets were found in the Auger FOV~\cite{the_pierre_auger_collaboration_a_aab_et_al_search_2020, Apollonio_ICRC_2025}. Also, no elongated patterns expected from coherent deflections were found in a principal-axis analyses. In a targeted search at Cen A's position, a 9-plet was found with a $p$-value of 6\%~\cite{Apollonio_ICRC_2025}. 
In another preliminary scan on Auger open data using a dedicated likelihood including parameters of the magnetic field~\cite{he_evidence_2024}, a multiplet is identified with a post-trial significance of $3.3\sigma$, including also one of the highest-energy events measured by Auger at $165\,\mathrm{EeV}$. Note that the multiplet itself is in the Cen A excess region, yet - considering coherent magnetic field deflections - the most probable source is identified to be the \textit{Sombrero galaxy}, an AGN at around $9\,\mathrm{Mpc}$ distance. Note that the likelihood to find multiplets will in general always be higher in flux excess regions (such as the Cen A region). This effect is not penalized for in~\cite{he_evidence_2024}. They also identify a multiplet pointing towards Cen A, albeit with a lower significance.
Another targeted search conducted by TA has found a $4.1\sigma$ indication for energy-ordered multiplets along the supergalactic plane~\cite{the_telescope_array_collaboration_r_u_abbasi_evidence_2020} that could be indicative of UHECRs diffusing from sources along that plane and hence roughly the LSS.

In the future, it will be very insightful to search for rigidity-ordered multiplets instead of just energy-ordered ones using the event-level mass estimators being developed by Auger~\cite{pierre_auger_collaboration_inference_2025} and TA~\cite{ prosekin_evaluation_2025}.
This is especially important because the UHECR mass composition changes so quickly with energy. 
A rigidity-ordered multiplet pointing unambiguously into the direction of a source like Cen A would be a clear indication of it being an actual UHECR source, and also the first direct observation of coherent magnetic field deflections. On the other hand, the reconstruction of the magnetic field indirectly from UHECR deflections might also become possible~\cite{golup_source_2009}, especially using machine-learning methods (see e.g.~\cite{wirtz_towards_2021, schulte_all-sky_2024}).

\subsection{Constraints from the highest-energy events} \label{sec:high_E_events}
At even higher energies than where the intermediate-scale anisotropies have started to arise, the event number decreases quickly. But, due to the shrinking propagation horizon with the energy, these events have to originate from relatively nearby sources so that correlation studies with nearby matter or auto-correlation may still be very informative. 
Before the start of Auger and TA, when the event statistic was still very limited, the arrival directions of UHECRs $>40\,\mathrm{EeV}$ seemed to exhibit clustering in doublets and triplets~\cite{takeda_small-scale_1999, uchihori_cluster_2000}, which was utilized to constrain properties of UHECR sources like the source number density, see e.g.~\cite{sigl_maximum_1997, dubovsky_statistics_2000, sigl_ultra-high_2004, singh_gamma-ray_2004, kachelriess_ultra-high_2005}. Using an early Auger data set from 2011, a limit of $n_s>6\times 10^{-6}\,\mathrm{Mpc}^{-3}$ was set on the source number density $>70\,\mathrm{EeV}$ - but that limit only applies if magnetic field deflections are $<30^\circ$. As visible in Fig.~\ref{fig:GMF_median_defl}, deflections can be a lot larger for a realistic mass composition heavier than nitrogen at $70\,\mathrm{EeV}$. No comparable analyses taking into account heavier mixed composition and updated GMF models are yet available. 

\begin{figure}[ht]
\centering
\includegraphics[width=.52\linewidth]{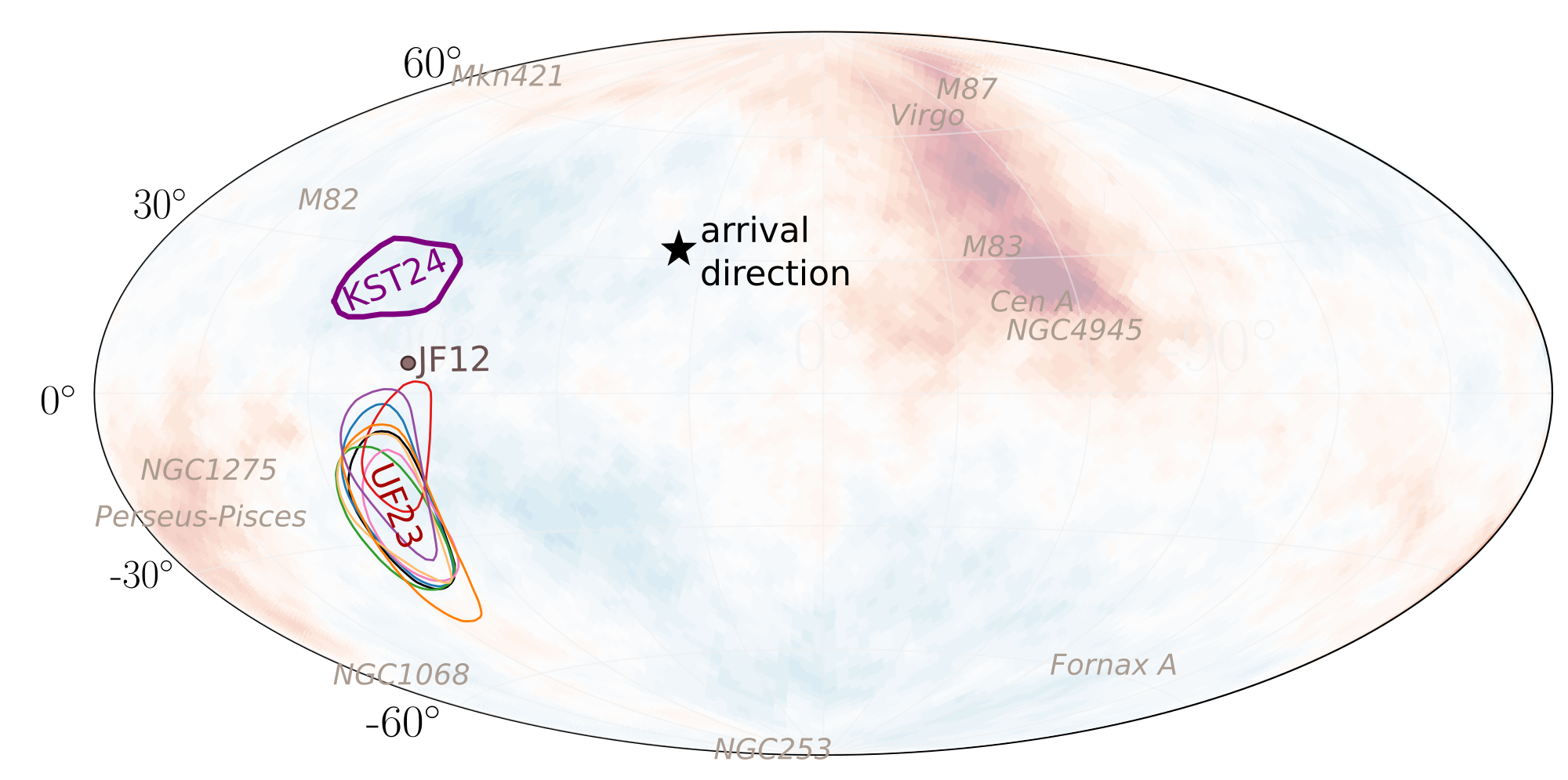}
\includegraphics[width=.47\linewidth]{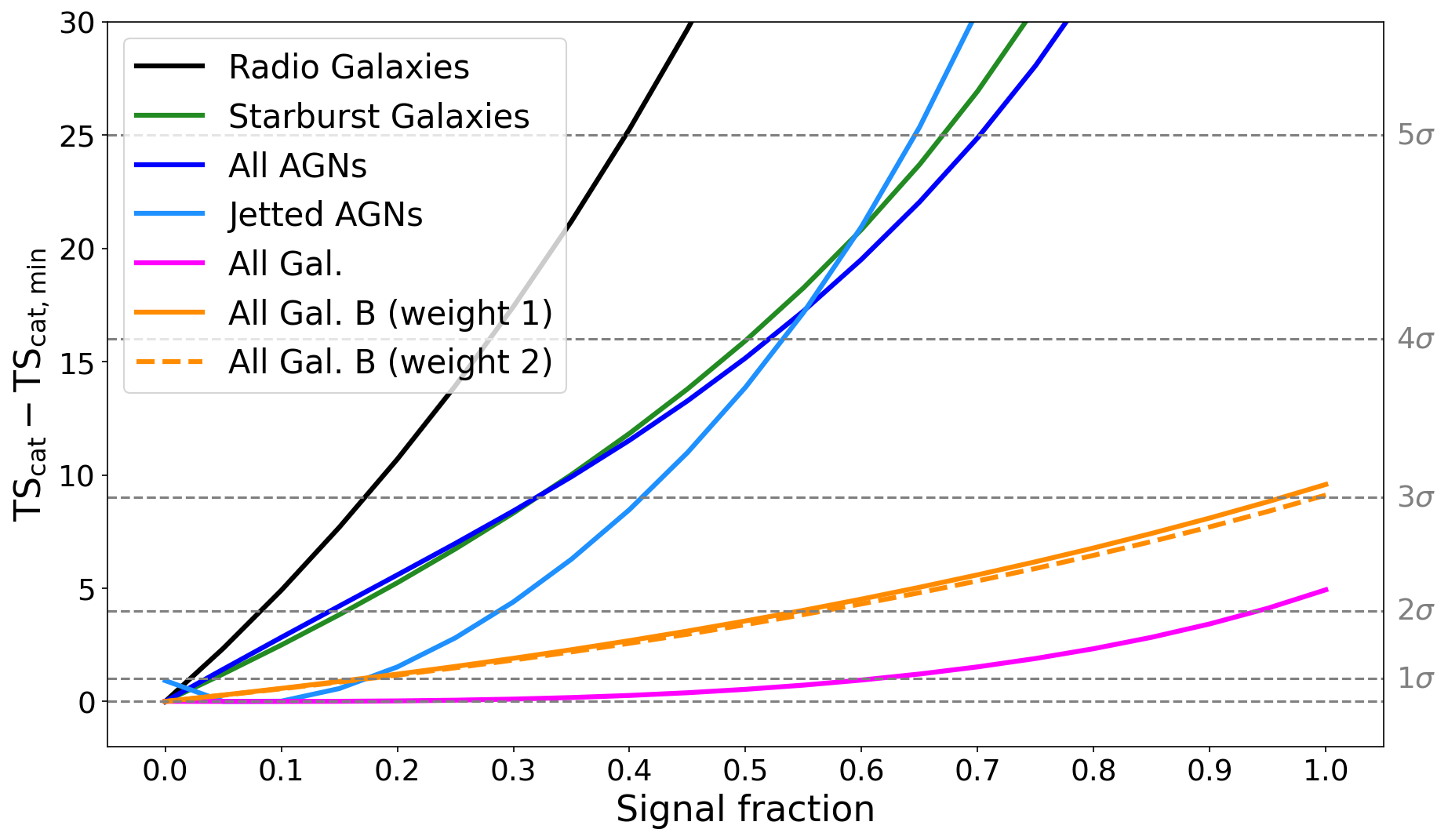}
\caption{\textit{Left:} Possible origins of the Amaterasu event~\cite{the_telescope_array_collaboration_r_u_abbasi_extremely_2023} in Galactic coordinates. The measured arrival direction is indicated by a black star. The backtracked possible origins for the \texttt{KST24} GMF model (including variations of the model parameters and energy within $1\sigma$ plus a $5^\circ$ turbulent blurring) is shown by a purple contour, taking the lower energy bound of $212\,\mathrm{EeV}$ and assuming an iron nucleus (from~\cite{korochkin_uhecr_2025}). The multiple colorful contours indicate the backtracked origin for the \texttt{UF23} GMF models with varying random fields, also assuming an iron nucleus and taking the lower energy bound (from~\cite{unger_where_2024}). For reference, the backtracked direction in the \texttt{JF12} model is also indicated for an iron nucleus (from~\cite{unger_where_2024}).
The expected flux of UHECRs from the LSS (from~\cite{bister_constraints_2024}, as in Fig.~\ref{fig:illumination}) is shown in the background, demonstrating that the backtracked directions point towards the local void (blue) instead of an overdense regions (red). The directions of several nearby source candidates are also indicated.
\textit{Right:} Test statistic for associating the 40 highest-energy Auger events with different source catalogs, assuming the \texttt{UF23} GMF models and a mixed composition, from~\cite{Bianciotto_ICRC_2025}. Only a $\sim10\%$ contribution from the Fermi-LAT catalog of jetted AGNs (light blue) fits slightly better than isotropy. A contribution $\gtrsim35\%$ from the van-Velzen catalog of radio galaxies, as well as $\gtrsim60\%$ from the Swift-BAT AGN catalog (dark blue), the Lunardini SBG catalog (green), and the Fermi-LAT catalog (light blue), are excluded at $5\sigma$. 
}
\label{fig:amaterasu}
\end{figure}

Recently, TA has measured an event with an exceptionally large energy of $E=244\pm29\,(\mathrm{stat})^{+51}_{-76}\,(\mathrm{syst})\,\mathrm{EeV}$ referred to as the "Amaterasu" event~\cite{the_telescope_array_collaboration_r_u_abbasi_extremely_2023}. By backtracking its arrival direction using different models of the GMF and assuming that the particle was originally an iron nucleus, it was found that Amaterasu most likely originated from a direction within the local void, and cannot be associated with any known local powerful galaxy or cluster~\cite{the_telescope_array_collaboration_r_u_abbasi_extremely_2023, unger_where_2024, korochkin_uhecr_2025, bourriche_beyond_2024}. This is shown in Fig.~\ref{fig:amaterasu} for different models of the GMF and associated uncertainties.
Due to its large energy, the source has to be close~\cite{kuznetsov_nearby_2024}, within $\sim(8-50)\,\mathrm{Mpc}$ depending on the exact event energy and assuming an iron nucleus at the source~\cite{unger_where_2024}. 
For all TA events $>100\,\mathrm{EeV}$ to be associated with the local matter distribution, either a very strong EGMF has to be present, or the composition has to be extremely heavy~\cite{telescope_array_collaboration_isotropy_2024}.
For Auger, a statistical analysis was performed to test the association between the events $>100\,\mathrm{EeV}$ and different source catalogs~\cite{Bianciotto_ICRC_2025}, as displayed in Fig.~\ref{fig:amaterasu} (\textit{right}). Not even a small contribution from any of the catalogs fits significantly better than isotropy. A contribution $\gtrsim60\%$ from the Lunardini catalog of starburst galaxies, the Swift-BAT AGN catalog, and the Fermi-Lat catalog of jetted AGNs is even excluded at $5\sigma$. Note that deflections by the EGMF are currently neglected in the analysis.

This non-association of the highest energy events with local source candidates or the LSS has sparked theories about the origin of Amaterasu and the highest energy events in general: next to a strong EGMF~\cite{mollerach_case_2024} or more exotic theories like beyond-standard-model physics or superheavy dark matter~\cite{lang_new_2024, sarmah_amaterasu_2024, frampton_amaterasu_2024}, it could simply be that the sources indeed lie in voids and are just not special or powerful enough to be a prominent source candidate. In that case, UHECRs could originate from transient events happening in these ordinary galaxies~\cite{unger_where_2024}, such as TDEs, young magnetars, or BNS mergers~\cite{farrar_binary_2025}. Another possibility is that UHECR generation anti-correlates with the electromagnetic source power, so that less prominent sources like the Sombrero galaxy (see sec.~\ref{sec:IS}) are responsible for UHECR production~\cite{anchordoqui_sombrero_2025}. For these lower-power sources, lower photon densities are expected, which may allow UHECRs to escape without heavy losses and hence without producing a strong $\gamma$-ray flux, see also~\cite{partenheimer_ultra-high-energy_2024}.

Another possibility is that the highest energy events are in fact heavier than iron~\cite{anchordoqui_unmasking_2000} and produced for example in r-processes happening in BNS mergers or collapsars~\cite{zhang_ultraheavy_2024, farrar_binary_2025, guo_binary_2025}. In that case, the propagation distance is still $\mathcal{O}(100\,\mathrm{Mpc})$ at $300\,\mathrm{EeV}$~\cite{zhang_ultraheavy_2024, andrade_dourado_towards_2025}, so that the sources could be further away. Additionally, deflections in the GMF would be larger for higher charge numbers. For the Amaterasu event, this means that sources on the supergalactic plane fall within the possible backtracked directions~\cite{zhang_ultraheavy_2024}.

\section{Conclusions and Outlook} \label{sec:conclusion}
Since the detection of UHECRs, their origin has remained a mystery. Thanks to the large increase in statistics of data by the Pierre Auger Observatory and Telescope Array, however, several characteristics of UHECR sources have been narrowed down over the last years. Quite a lot of the findings were unexpected, such as the intermediate to heavy mass composition at the highest energies. By adapting models to best describe the measured energy spectrum, shower depth distributions, and lately also the arrival directions, the following characteristics of UHECR sources have been identified as most likely in the energy range above the ankle ($\sim(8-32)\,\mathrm{EeV})$:
\begin{itemize}
    \vspace{-0.3cm}\item They are extragalactic and evolve with a source evolution weaker than $\propto(1+z)^5$~\cite{the_pierre_auger_collaboration_a_aab_combined_2017, the_pierre_auger_collaboration_a_abdul_halim_constraining_2023, the_pierre_auger_collaboration_a_abdul_halim_et_al_constraining_2024, alves_batista_cosmogenic_2019} (see sec.~\ref{sec:CF}).
    \vspace{-0.4cm}\item They emit (possibly after in-source interactions or magnetic confinement~\cite{unger_origin_2015}) a relatively hard spectrum and mixed composition~\cite{the_pierre_auger_collaboration_a_aab_combined_2017, the_pierre_auger_collaboration_a_abdul_halim_constraining_2023, the_pierre_auger_collaboration_a_abdul_halim_et_al_constraining_2024} where the maximum rigidity does not differ much between sources~\cite{ehlert_curious_2023, Luce_ICRC_2025} (see sec.~\ref{sec:CF}).
    \vspace{-0.4cm}\item They at least roughly follow the large-scale structure~\cite{waxman_signature_1997, cuoco_footprint_2006, harari_anisotropies_2015, tinyakov_full_2015, globus_extragalactic_2017, di_matteo_how_2018, globus_cosmic_2019, ding_imprint_2021, allard_what_2022, bister_constraints_2024, bister_large-scale_2024, abdul_halim_large-scale_2024} and have a relatively large density $n_s\gtrsim10^{-4}\,\mathrm{Mpc}^{-3}$ (see sec.~\ref{sec:LS})~\cite{allard_what_2022, bister_constraints_2024, bister_large-scale_2024, abdul_halim_large-scale_2024}.
    \vspace{-0.2cm}
\end{itemize}
Especially the last two points have sparked debates about which kind of source class could be on one hand so common, and on the other hand so identical~\cite{farrar_binary_2025}. Also, further modeling is needed to gain information about the exact relevant energy range where the above listed findings apply, i.e. if the same source class could be responsible for the dipole, intermediate-scale anisotropies, and the highest energy events. The sources of the smaller-scale anisotropies are still heavily under debate, e.g. the source behind the flux excess in the Cen A region seen by Auger. The overdensity could be caused for example by Cen A itself~\cite{farrar_deducing_2000, gorbunov_comment_2007, taylor_uhecr_2023}, by the nearby starburst galaxy NGC4945~\cite{the_pierre_auger_collaboration_a_aab_et_al_indication_2018, the_pierre_auger_collaboration_a_abdul_halim_et_al_constraining_2024}, or by deflected events from the Virgo cluster~\cite{ding_imprint_2021, allard_what_2024} or the Sombrero galaxy~\cite{he_evidence_2024, anchordoqui_sombrero_2025} (see sec.~\ref{sec:IS}). Another pressing question has emerged regarding the highest energy events, which cannot easily be associated with nearby prominent source candidates or the extragalactic matter distribution~\cite{the_telescope_array_collaboration_r_u_abbasi_extremely_2023, unger_where_2024, korochkin_uhecr_2025, unger_galactic_2025}. If these events are produced by the same source class as dominant above the ankle, the sources are likely of transient nature - such as binary neutron star mergers~\cite{farrar_binary_2025} - and the mass composition at the highest energies is likely heavier than iron~\cite{anchordoqui_unmasking_2000, zhang_ultraheavy_2024} (see sec.~\ref{sec:high_E_events}). 

Some of these questions could be answered in the near future by using mass estimators from the Auger and TA surface detectors~\cite{pierre_auger_collaboration_inference_2025, prosekin_evaluation_2025}. Combining event-by-event charge information at the highest energies with energy and arrival direction measurements will provide powerful constraints on UHECR sources. This could be done for example by extending combined fits of models to multiple observables such as~\cite{the_pierre_auger_collaboration_a_abdul_halim_et_al_constraining_2024}, by investigating the difference between heavy and light components in large- and intermediate-scale anisotropies~\cite{martins_prospects_2025, apollonio_amplifying_2025}, or by searching for rigidity orderings near source candidates.
Additionally, multi-messenger analyses can be a powerful probe of UHECR sources. First hints about UHECR sources from neutrinos are for example the observation of neutrinos from the nearby starburst galaxy NGC1068~\cite{collaboration_evidence_2022}, a neutrino associated with a tidal disruption event~\cite{stein_tidal_2021}, or the non-observation of neutrinos from gamma-ray bursts~\cite{abbasi_searches_2022}.

When comparing the findings listed in this review to the open questions raised in~\cite{alves_batista_open_2019}, it can be seen that several of those questions have been at least partially answered in the last six years.
Yet, the next generation of UHECR experiments such as GCOS~\cite{ahlers_ideas_2025} or POEMMA~\cite{collaboration_poemma_2021} is needed to provide a further increase in statistics and precision of mass indicators, especially at the highest energies, to answer remaining questions. Additionally, further improvements in modeling have to be made, e.g. by even better models of the Galactic and extragalactic magnetic field.

\paragraph{Acknowledgements:} I thank Jonathan Biteau, Glennys Farrar, Foteini Oikonomou, Paul Simeon, Michael Unger, and Walter Winter - as well as all my colleagues from the Pierre Auger Collaboration - for fruitful discussions, and Silvia Mollerach and Armando di Matteo for comments on this manuscript. I also thank Alexander Korochkin and Michael Unger for providing their GMF models.

\begin{multicols}{2}
\setlength{\bibsep}{0.0pt}
\footnotesize
\bibliographystyle{JHEP}
\bibliography{references}
\end{multicols}

\end{document}